\documentclass[usenatbib]{mn2e}
\usepackage{graphicx, amssymb, aas_macros, bm}
\newcommand{\vmax}{V_{\rm{max}}}
\newcommand{\rmax}{R_{\rm{max}}}

\newcommand{\msub}{M_{\rm{sub}}}

\newcommand{\mvir}{M_{\rm{vir}}}
\newcommand{\mtwo}{M_{200}}
\newcommand{\mtwom}{M_{200 m}}
\newcommand{\rtwo}{R_{200}}
\newcommand{\rtwom}{R_{200 m}}

\newcommand{\rvir}{R_{\rm{vir}}}
\newcommand{\vvir}{V_{\rm{vir}}}
\newcommand{\dd}{{\rm d}}
\newcommand{\mstar}{M_{\star}}
\newcommand{\msun}{M_{\odot}}
\newcommand{\hmsun}{h^{-1} \, M_{\odot}}
\newcommand{\hmpc}{h^{-1} \, {\rm Mpc}}
\newcommand{\hkpc}{h^{-1} \, {\rm kpc}}

\newcommand{\kms}{{\rm km \, s}^{-1}}
\newcommand{\millen}{MS-II}
\newcommand{\zacc}{z_{\rm acc}}
\newcommand{\vacc}{V_{\rm acc}}
\newcommand{\macc}{M_{\rm acc}}
\newcommand{\muacc}{\mu_{\rm acc}}
\newcommand{\cvf}{\langle N(>\nu) \rangle}
\newcommand{\cmf}{\langle N(>\mu) \rangle}
\newcommand{\cmfacc}{\langle N(>\mu_{\rm acc}) \rangle}
\newcommand{\vmaxo}{V_{\rm{max},0}}
\newcommand{\msubo}{M_{\rm{sub},0}} 
\newcommand{\mviro}{M_{\rm{vir},0}} 
\newcommand{\mmerge}{M_{\rm merge}}
\newcommand{\lcdm}{$\Lambda$CDM}
\newcommand{\muone}{\widetilde{\mu_1}}

\title[Statistics of Milky Way-mass halos]
{
There's no place like home?  Statistics of Milky Way-mass dark matter halos
}

\author[M.~Boylan-Kolchin et al.]{
  $\!\!$Michael~Boylan-Kolchin$^1$\thanks{$\!\!$e-mail:
    mrbk@mpa-garching.mpg.de}, Volker~Springel$^1$, Simon~D.~M.~White$^1$
 and \newauthor Adrian~Jenkins$^2$\\
  $\!\!^1$Max-Planck-Institut f\"{u}r Astrophysik, Karl-Schwarzschild-Str. 1,
  85748 Garching, Germany\\
  $\!\!^2$Institute for Computational Cosmology, Department of Physics,
  University of Durham, South Road, Durham DH1 3LE, UK }
\begin{document}

 \pagerange{\pageref{firstpage}--\pageref{lastpage}} 
 \pubyear{2010}

\maketitle

\label{firstpage}
\begin{abstract}
  We present an analysis of the distribution of structural properties for Milky
  Way-mass halos in the Millennium-II Simulation (MS-II).  This simulation of
  structure formation within the standard \lcdm\ cosmology contains thousands
  of Milky Way-mass halos and has sufficient resolution to properly resolve
  many subhalos per host.  It thus provides a major improvement in the
  statistical power available to explore the distribution of internal structure
  for halos of this mass.  In addition, the MS-II contains lower resolution
  versions of the Aquarius Project halos, allowing us to compare our results to
  simulations of six halos at a much higher resolution.  We study the
  distributions of mass assembly histories, of subhalo mass functions and
  accretion times, and of merger and stripping histories for subhalos capable
  of impacting disks at the centers of halos.  We show that subhalo abundances
  are {\it not} well-described by Poisson statistics at low mass, but rather
  are dominated by intrinsic scatter.  Using the masses of subhalos at infall
  and the abundance-matching assumption, there is less than a 10\% chance that
  a Milky Way halo with $\mvir =10^{12}\,\msun$ will host two galaxies as
  bright as the Magellanic Clouds.  This probability rises to $\sim 25\%$ for a
  halo with $\mvir=2.5 \times 10^{12} \,\msun$.  The statistics relevant for
  disk heating are very sensitive to the mass range that is considered
  relevant.  Mergers with infall mass : redshift zero virial mass greater than
  1:30 could well impact a central galactic disk and are a near inevitability
  since $z=2$, whereas only half of all halos have had a merger with infall
  mass : redshift zero virial mass greater than 1:10 over this same period.
\end{abstract}
\begin{keywords}
 methods: $N$-body simulations -- cosmology: theory -- galaxies: halos
\end{keywords}

\section{Introduction} 
Our understanding of galaxy formation is strongly influenced by our own Galaxy.
Several of the apparent points of tension between observations and the
predictions of the now-standard $\Lambda$ Cold Dark Matter (\lcdm) structure
formation model -- for example, the abundance of dark matter subhalos in
comparison to the relative paucity of observed Local Group satellites and the
existence of the thin Galactic disk in light of the ubiquity of dark matter
halo mergers -- are most clearly demonstrated from Milky Way data.  While it is
clearly an excellent testing ground for a wide array of phenomena related to
galaxy formation, the Milky Way (MW) is nonetheless a single object and may not
be representative in its properties.  This is particularly true with respect to
infrequent, stochastic phenomena such as major mergers: while \lcdm\ robustly
predicts the mean number of mergers onto halos of a given mass, the
halo-to-halo scatter is large.

The situation is similar for cosmological re-simulations of individual MW-mass
dark matter halos.  In the quest to understand the Milky Way's formation,
numericists have been simulating individual MW-mass dark matter halos at
increasingly high resolution.  By investing substantial computational
resources, the most recent generation has reached one billion particles per
halo, with particle masses $\sim\! 1000 \,\msun$ \citep{diemand2008,
  springel2008, stadel2009}.  Only a small number of such simulations can be
performed and there is no guarantee that the simulated halos will be
``typical.''

In order to explore whether various characteristics of the Milky Way (e.g., 
its satellite galaxy population, thin disk, or merger history) are
consistent with \lcdm\ expectations, we must understand the statistical
properties predicted for MW-mass halos.  The advent of modern galaxy surveys
has enabled such an undertaking observationally: the Sloan Digital Sky Survey
(SDSS; \citealt{york2000}) has provided a huge sample of galaxies with stellar
masses similar to that of the MW, although only mean halo properties averaged
over large numbers of similar galaxies can readily be measured
\citep{mandelbaum2006b, more2009}.  Predicting the formation histories and
redshift zero properties of the halos of a representative population of
MW-mass galaxies requires numerical simulations that combine large volume with
high mass resolution.  Resolving the internal structure of $\sim$ ten subhalos
per host halo requires approximately one hundred thousand particles per host,
which is not trivial when combined with the goal of obtaining a statistical
sample of MW-mass halos.  This would, for example, require $10^{12}$ particles in
the volume probed by the Millennium Simulation (MS; \citealt{springel2005b}), a
100-fold increase over the particle number actually used.

The only computationally feasible approach at present is to use a smaller
simulation volume combined with higher mass resolution.  In this paper, we use
the Millennium-II Simulation (MS-II; \citealt{boylan-kolchin2009}), which was
{\it designed} to meet these requirements.  It allows us to investigate statistical
properties of MW-mass halos, including their assembly histories and their
subhalo content.  Throughout this work, we also compare results for our
representative sample to the six halos from the Aquarius simulation series of
\citet{springel2008}.  These cosmological resimulations of Milky Way-mass dark
matter halos have approximately one thousand times better mass resolution than
the \millen\ (see Section~\ref{subsec:aquarius}).

The basic properties of the \millen\ and of the sample of MW-mass dark matter
halos we investigate are given in Section~\ref{sec:simulations}, together with
a convergence study for subhalos.  Section~\ref{sec:mainHalos} discusses the
assembly history of MW-mass halos in terms both of their virial mass and of
their central gravitational potential, and relates these to their $z=0$
concentrations and angular momenta.  Section~\ref{sec:subhalos} investigates
the subhalo population of MW-mass halos at $z=0$, including the subhalo mass
function and whether its scatter is Poissonian, the distribution of masses for
the most massive subhalo in each host, and typical accretion times for subhalos
as a function of their current mass.  In Section~\ref{sec:mergers}, we explore
the merger histories of MW-mass halos, focusing on mergers that are likely to
result in galaxy-galaxy mergers.  Section~\ref{sec:discussion} contains a
discussion of our results and their implications, while
Section~\ref{sec:conclusions} gives our conclusions.  Some properties of
satellite subhalos are quantified in the Appendix.  Throughout this paper, all
logarithms without a specified base are natural logarithms.

\section{Numerical Simulations}
\label{sec:simulations}
\subsection{The Millennium-II Simulation}
\label{subsec:msII}
The Millennium-II Simulation is a very large cosmological $N$-body simulation
of structure formation in the \lcdm\ cosmology.  The simulation is described
fully in \citet{boylan-kolchin2009}.  For completeness, we review some of its
salient features here.

The \millen\ follows $2160^3 \approx 10$ billion particles,
each of mass $m_p=6.885 \times 10^6 \, \hmsun$, from redshift $z=127$ to $z=0$
in a periodic cube with side length $L=100 \, \hmpc$.  The Plummer-equivalent
force softening of the simulation is $\epsilon=1 \, \hkpc$ and was kept
constant in comoving units for the duration of the simulation.

The cosmological parameters used in the \millen\ are identical to those adopted
for the Millennium Simulation and the Aquarius simulations:
\begin{eqnarray}
 \label{eq:cosmo_params}
 && \Omega_{\rm tot} = \! 1.0, \; \Omega_m = \!0.25, \; \Omega_b=0.045, \; 
 \Omega_{\Lambda}=0.75, \nonumber \\
 && h = 0.73, \;\sigma_8=0.9, \; n_s=1\,,
\end{eqnarray}
where $h$ is the Hubble constant at redshift zero in units of $100 \, \kms\,
{\rm Mpc}^{-1}$, $\sigma_8$ is the rms amplitude of linear mass fluctuations in
$8 \,\hmpc$ spheres at $z=0$, and $n_s$ is the spectral index of the primordial
power spectrum.  While the values of these parameters were originally chosen
(for the MS) for consistency with results from the combination of WMAP 1-year
and 2dF Galaxy Redshift Survey data \citep{spergel2003, colless2001}, some are
now only marginally consistent with more recent analyses.  In particular,
$\sigma_8 = 0.812$ and $n_s = 0.96$ are now the preferred values based on WMAP
5-year results combined with baryon acoustic oscillation and Type Ia supernova
data \citep{komatsu2009}.  For the most part, these differences are unimportant
for the results presented in this paper: the abundance of $10^{12}\,\hmsun$
halos in the WMAP5 and Millennium-II cosmologies differs by less than 7\% at
$z=0$, for example.  A discussion about possible effects of varying the assumed
cosmology is contained in Sec.~\ref{subsec:cosmo_params}.

In addition to the raw simulation output of six phase-space coordinates and one
identification number per particle, stored at 68 outputs spaced according to
equation~2 of \citet{boylan-kolchin2009}, additional information about dark
matter structures and merging histories in the \millen\ was computed later and
saved.\footnote{Dark matter halo catalogs (both FOF and {\tt
    SUBFIND}) and
  merger trees from the \millen\ are publicly available at\\
  http://www.mpa-garching.mpg.de/galform/millennium-II}

Friends-of-Friends (FOF) groups were identified on-the-fly during the
simulation using a linking length of $b=0.2$.  All groups at each time with at
least 20 particles were stored in halo catalogs.  FOF groups were searched for
bound substructure using the {\tt SUBFIND} algorithm \citep{springel2001a}.
All substructures containing at least 20 bound particles were deemed to be
physical subhalos and were stored in subhalo catalogs at each snapshot.  Each
subhalo in the \millen\ was assigned a mass $\msub$ equal to the sum of its
constituent particles; this is the mass for subhalos we use throughout this
paper.  All subhalos also have a maximum circular velocity $\vmax$, defined via
\begin{equation}
  \label{eq:vmax}
  \vmax^2 \equiv {\rm max} \left(\frac{G \msub(<r)}{r} \right) \,.
\end{equation}
The main component of a FOF halo is generally a dominant subhalo\footnote{This
  is sometimes referred to as the ``central'' subhalo and corresponds closely to
  the standard picture of a smooth, virialized dark matter halo } whose
particles make up most of the bound component of the FOF group.  FOF
halos can also contain non-dominant subhalos (``satellites''); these typically
contain in total only a small fraction of the mass of the dominant subhalo,
even in the highest resolution simulations currently possible \citep{gao2004b,
  springel2008, diemand2008, stadel2009}.  

Each FOF halo containing at least one subhalo has an associated virial mass and
virial radius, defined as the mass and radius of a sphere containing an average
density $\Delta$ times the critical value $\rho_c(z)$
Throughout this paper, we use $\Delta = \Delta_{\rm vir}$, which is derived
from the spherical top-hat collapse model (e.g., \citealt{gunn1972,
  bryan1998}); we denote the virial mass $\mvir$ and the virial radius $\rvir$.
Other common choices include $\Delta=200$ (corresponding to $\mtwo$ and
$\rtwo$) and $\Delta=200 \,\Omega_m(z)$ [corresponding to $\mtwom$ and
$\rtwom$]. At $z=0$, $\Delta_{\rm vir} \approx 94$ for the cosmology in
equation~(\ref{eq:cosmo_params}) and $\mtwo < \mvir < \mtwom$ for any given
halo.  For dominant subhalos, $\msub$ is comparable to, but usually slightly
less than, $\mvir$.

Merger trees were constructed at the subhalo level by requiring each subhalo to
have at most one unique descendant.  Descendants were determined by considering
all subhalos at snapshot $S_{N+1}$ containing at least one particle from a
given subhalo at snapshot $S_{N}$ and computing a weighted score based on the
particles' binding energies from $S_N$.  The descendant-finding algorithm also
checks possible descendants at $S_{N+2}$ to account for cases where a subhalo
temporarily disappears from the catalogs, usually because it passes very near
another, more massive subhalo.  Note that the merger trees are {\it not}
constructed to explicitly follow mergers of FOF groups but rather to follow
mergers of individual subhalos; the merger of two subhalos is often delayed
relative to the merger of their original host FOF groups by several dynamical
times.  Such subhalo mergers correspond much more closely to the intuitive
notion of disjoint objects merging to form a monolithic descendant than do
mergers of FOF groups.

\subsection{The Aquarius Simulations}
\label{subsec:aquarius}
The Aquarius Project \citep{springel2008} is a suite of cosmological
resimulations of Milky Way-mass dark matter halos.  Six halos, denoted Aq-A
through Aq-F, were selected at $z=0$ from a cosmological simulation known as
hMS \citep{gao2008}.  The hMS is a lower resolution version of the \millen: it
follows $900^3$ particles in the same volume, with the same cosmological
parameters, and with the same amplitudes and phases\footnote{The amplitudes and
  phases are identical for all perturbations with wavenumber smaller than the
  Nyquist frequency of the hMS; higher wavenumber modes in the \millen\ were
  randomly sampled from the same underlying power spectrum} for the initial
power spectrum as the \millen.  The halos were selected randomly from the full
set of Milky Way-mass halos in the hMS simulation subject only to a weak
isolation criterion: the nearest halo with mass greater than half that of the
candidate halo was required to be at least $1 \,\hmpc$ from the candidate halo
at $z=0$.  This isolation criterion is not particularly restrictive: only 22\%
of the halos from the mass-selected sample in the \millen\ described below fail
to meet this requirement.  Once selected, the halos were re-simulated at up to
five different levels of resolution.  The highest level, which used a particle
mass of $1.2 \times 10^{3} \,\hmsun$ and a force softening of $14 \,h^{-1}$
parsecs, led to over 1.5 billion particles within $\rtwom$ at $z=0$.  This
resolution was used for only one of the six halos but all six were resimulated
at the next resolution level, for which $m_p$ is at most $10^{4} \,\hmsun$.
These are the ``level 2'' simulations that we compare to the \millen\ halos in
this paper.  Subhalo catalogs and merger trees analogous to those from the
\millen\ were also created for the Aquarius simulations.  Further details of
the Aquarius Project are presented in \citet{springel2008}.

\subsection{Milky Way-mass halos}
Recent estimates of the mass of the Galaxy's dark matter halo range from $1-3
\times 10^{12}\,\msun$ \citep{wilkinson1999, sakamoto2003, battaglia2005,
  dehnen2006, xue2008, li2008a}.  To bracket these estimated values
of $M_{\rm vir, MW}$ and to assess any possible trends with halo mass, we
select all halos with $10^{11.5 } \le \mvir / [\hmsun] \le 10^{12.5}$ from the
redshift zero output of the MS-II.  The resulting sample of 7642 total ``Milky
Way-mass'' halos forms the basis of the analysis in this paper.  With the mass
resolution of the \millen\ -- $m_p=6.885 \times 10^6 \, \hmsun$  --
this selection criterion means that the MW-mass halos considered here are
represented by 50,000-500,000 particles.  The typical virial radius of these
halos is $\rvir \approx 200 \, \hkpc$, which is 200 times larger than the
Plummer-equivalent force softening of the MS-II.

An alternative to selecting halos by mass is instead to select by $\vmax$,
which is more observationally accessible.  Such a selection is not without
complications, however, as the presence of a disk and bulge is likely to
significantly alter $\vmax$ in real galaxies.  In the models of
\citet{klypin2002}, for example, the adiabatic response of the dark matter halo
to the formation of the MW causes an increase in $\vmax$ of between 19\% and
100\%.  The virial mass of the halo should not be modified by these effects.

\subsection{Subhalos}
\begin{figure}
 \centering
 \includegraphics[scale=0.55, viewport=0 0 440 780]{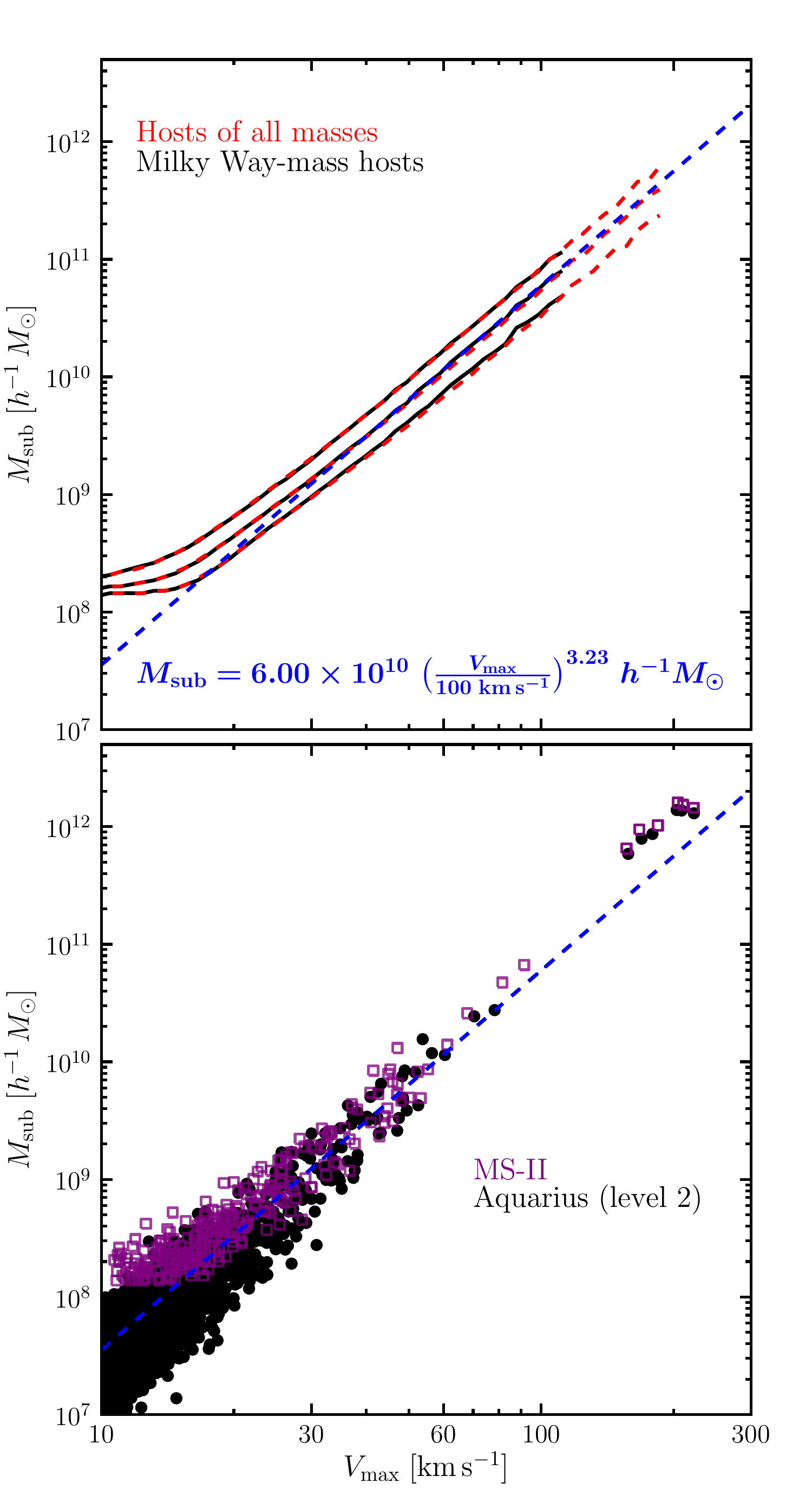}
 \caption{ A comparison of various $\vmax-\msub$ relations for satellite
   subhalos with 
   $d_{\rm sub} < \rvir$ at redshift zero.  {\it Top:} the median and $\pm
   1 \sigma$ $\vmaxo-\msubo$ relation for subhalos of Milky Way-mass halos
   (black) and for subhalos of all halos (red) from the \millen.  The
   best power law fit to the median relation is plotted in blue.  {\it Bottom:} A direct
   comparison of $\vmaxo-\msubo$ for the six level 2 Aquarius halos (filled
   black circles) and for the corresponding six halos in the \millen\ (open magenta
   squares); the best-fit relation from the upper panel is also
   plotted as a blue dashed line.  The two data sets agree well for $\vmax \ga 25
   \,\kms$, while the \millen\ points lie systematically above the Aquarius
   points at lower $\vmax$ due to mass resolution effects.  The six central halos are
   also included in the bottom panel (upper right) and are offset from the
   satellite subhalo relation because they are not affected by tidal
   stripping.  
  \label{fig:massVsVmax}
}
\end{figure}
Halo substructure is a direct result of the hierarchical nature of halo
formation in the \lcdm\ cosmology.  Properties of subhalos such as their
abundance, spatial distribution, and internal structure can provide an
important link between simulations and observations, since (at least some)
subhalos are expected to be the hosts of satellite galaxies.  One should
consider only subhalos that are reliably resolved, however, as subhalo
structure can be modified by numerical effects at low resolution.  The upper
panel of Fig.~\ref{fig:massVsVmax} shows the relationship between $\vmax$ at
$z=0$ (denoted $\vmaxo$) and $\msub$ at $z=0$ ($\msubo$) for all satellite
subhalos in our MW-mass halo sample that lie within distance $\rvir$ from the
center of their host halos (black lines); the three lines show the $16^{\rm
  th}$, $50^{\rm th}$, and $84^{\rm th}$ percentiles of the distribution.  Also
plotted is the same relation for all host halos in the \millen\ (red dashed
lines).  The Milky Way sample agrees very well with the full simulation sample.

The upper panel of Fig.~\ref{fig:massVsVmax} also shows the best power law
$\vmaxo-\msubo$ relation fitted over the range $40 \le \vmaxo \le 80\,\kms$
(blue dashed curve) of $\msubo=6 \times 10^{10} \,(\vmaxo/100
\,\kms)^{3.23}\,\hmsun$.  The relation follows the power law fit closely for
$\vmax \ga 25 \, \kms$, while there is a strong break at low $\vmax$.  This is
due to mass resolution, as is confirmed by the lower panel of
Fig.~\ref{fig:massVsVmax}.  This panel shows the same relation but now for all
six of the level 2 Aquarius halos (black circles) and for the corresponding six
halos in the \millen\ (open magenta squares).  While the level 2 Aquarius data
follow the power-law relation down to the smallest $\vmax$ plotted
[$10\,\kms$], the \millen\ data start to differ systematically at $\vmax
\approx 25 \,\kms$.  Note that the host halos follow a relationship with a
similar slope but with larger mass at fixed $\vmax$.  This is in part because
central subhalos are not affected by tidal stripping.  In addition, the way
{\tt SUBFIND} assigns particles to bound structures, starting with all
particles bound to the dominant subhalo, contributes to the offset.  Given the
results of Fig.~\ref{fig:massVsVmax}, we adopt 150 particles ($\msub=1.03
\times 10^{9} \,\hmsun$), corresponding to $\vmaxo=28.3\,\kms$, as the minimum
number of particles for which we consider a subhalo well enough resolved to
estimate $\vmax$.  This corresponds to a cut where the $\vmax-\msub$ relation
deviates by about 10\% from the power law fit in the upper panel of
Fig.~\ref{fig:massVsVmax}.

Subhalos in high resolution simulations can lose substantial mass as they are
tidally stripped and gravitationally shocked within the potential of their
hosts (e.g., \citealt{tormen1998, klypin1999, ghigna2000, hayashi2003,
  kravtsov2004, gao2004b, boylan-kolchin2009}).  The putative galaxies within
these subhalos should be much more resilient to stripping and disruption both
because they sit in the densest regions of dark matter halos and because the
condensation of baryons to the centers of dark matter halos deepens the
gravitational potential \citep{white1978a}.

This differential stripping, affecting dark matter more strongly than stellar
components, means that the relationship between the stellar content and the
instantaneous dark matter properties of a subhalo is generally very
complicated.  It is much more likely that the mass (or $\vmax$) a subhalo had
when it was still an independent halo (i.e., a dominant subhalo in a FOF group)
can be related to its stellar content in a straightforward manner.  Indeed,
several studies have shown that many features of the observed galaxy
distribution can be reproduced in $N$-body simulations simply by assuming that
the stellar mass of a galaxy is a monotonically increasing function of the
maximum dark matter mass ever attained by the subhalo that surrounds it.  This
``abundance matching'' assumption \citep{vale2004} can reproduce observed
galaxy clustering and its evolution with redshift \citep{conroy2006, brown2008,
  conroy2009, wetzel2010, guo2010}.

Since it is useful to know when a given subhalo first became a satellite and
what its mass was at that time, we trace each subhalo back in time and find the
point at which its bound mass (determined by {\tt SUBFIND}) reached a maximum.
We denote this redshift by $\zacc$ and the corresponding subhalo mass and
maximum circular velocity as $M(\zacc) \equiv \macc$ and $\vmax (\zacc) \equiv
\vacc$.  The subscript ``acc'' is used because subhalos usually reach their
maximum mass immediately before being accreted into a more massive object; in
this sense, $\zacc$ represents an accretion redshift.\footnote{Even though the
  criterion defining $\zacc$ is that the subhalo's bound mass is maximized over
  its history, we avoid the subscript ``max'' so as not to cause any confusion
  with the peak in the circular velocity curve $\vmax$, which is defined at all
  redshifts for which a subhalo can be identified} This allows us to study
quantities such as the distribution of {$\zacc$} and the average mass loss of
satellite subhalos.  Relationships among $\macc$, $\vacc$, and $\vmaxo$ for
subhalos are presented in the Appendix.

\section{Main Halos}
\label{sec:mainHalos}
In this section, we study some properties of the main halos themselves.
Sec.~\ref{subsec:mah} explores the assembly histories of Milky Way mass
halos, both in terms of $\mvir$ and $\vmax$.  In Sec.~\ref{subsec:conc}, we
investigate the distribution of halo concentrations.  The focus of
Sec.~\ref{subsec:spin} is on halo spin parameters.  Throughout this section, we
compare properties of the six Aquarius halos to properties of the full \millen\
sample of ``Milky Way'' halos.

\subsection{Halo Assembly}
\label{subsec:mah}
\begin{figure}
 \centering
 \includegraphics[scale=0.55, viewport=0 0 440 780]{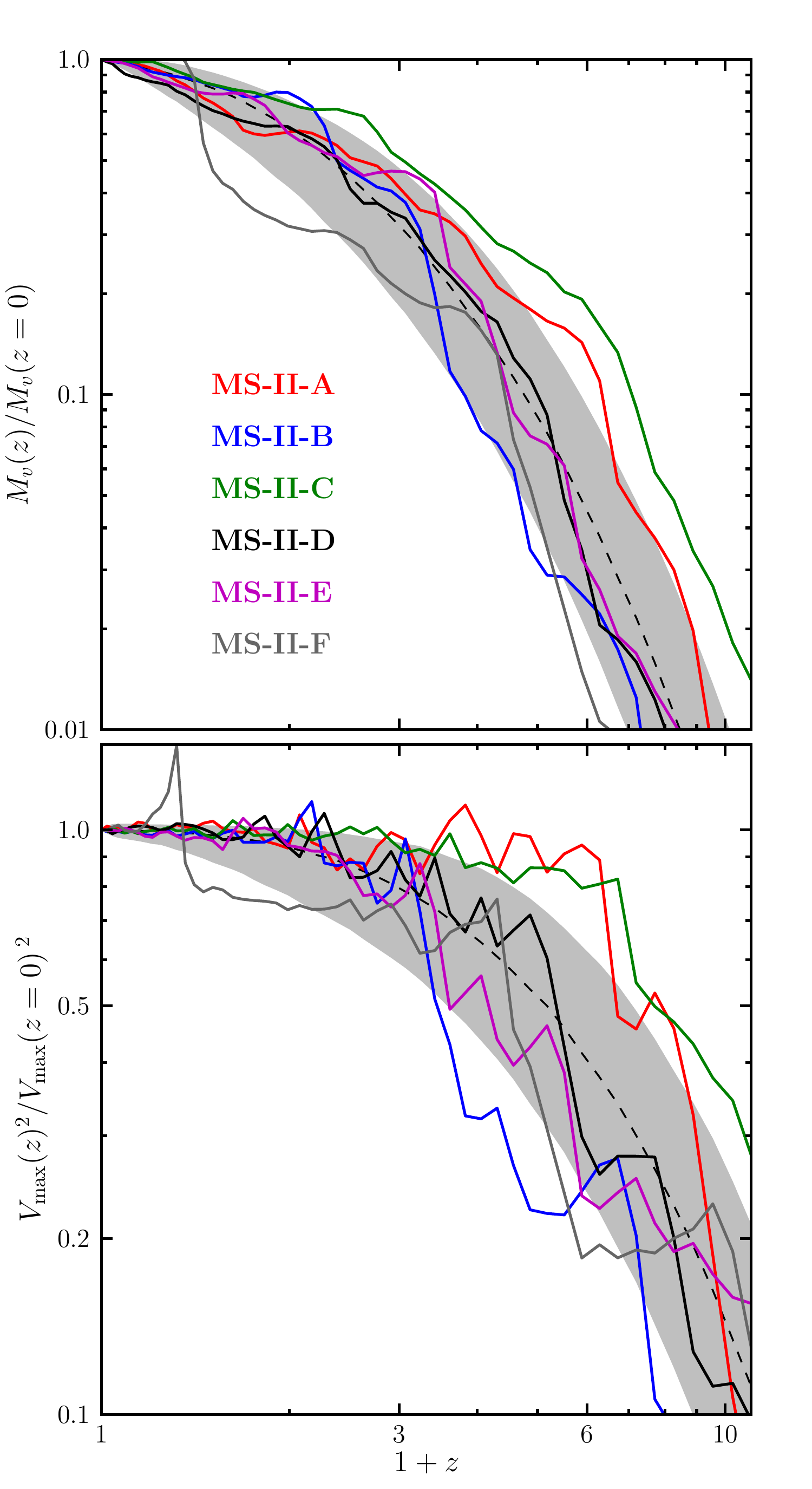}
 \caption{
   {\it Top}: Mass assembly histories of Milky Way-mass halos.  The median MAH
   is shown  as a dashed curve, while the gray shaded region contains 68\% of the
   distribution.  Also plotted are the MAHs for the \millen\ versions of the six
   Aquarius halos.  The Aquarius halos provide a fairly representative sample
   of 
   MAHs; halos A and C have quiescent merger histories, while halo F has a
   recent major merger.   {\it Bottom}: $\vmax$ accretion histories from the
   \millen.  As in the upper panel, the median relation is plotted as a dashed
   curve while the $\pm 1\,\sigma$ region is shown as the gray shaded region.
   The earliest-forming Aquarius halos in terms  of MAHs --  halos A and C --
   also assemble earliest as judged by $\vmax^2$, with $\vmax^2$ changing 
   by less than 25\% from $z=5$ to $z=0$.  Halo F, on the other hand, undergoes
   a major merger at $z \approx 0.4$ that changes $\vmax^2$ by 25\%.
   \label{fig:mah}
 }
\end{figure}
The mass assembly histories (MAHs) of dark matter halos have been studied
extensively via $N$-body simulations and extended Press-Schechter
(\citeyear{press1974}) theory \citep{lacey1993, lacey1994, wechsler2002,
  van-den-bosch2002, zhao2003, zhao2003a, tasitsiomi2004, cohn2005,
  neistein2006, li2008, zhao2009, mcbride2009}.  As a direct result of this
work, many basic properties of MAHs have been established.  For example,
\citet{wechsler2002} found that MAHs of individual halos follow a one-parameter
family $M(z)=M_0 \,e^{-\kappa\,z}$.  The parameter $\kappa$ is directly related
to the typical formation time for halos of mass $M_0$ and is an increasing
function of mass.  \citet{tasitsiomi2004} showed that this exponential form was
unable to fit individual cluster-mass MAHs in some cases and that a more
general function, $M(z)=M_0 \,(1+z)^{\eta}\, e^{-\kappa\,z}$, provided a
better match on a case-by-case basis.  \citet{mcbride2009} confirmed and
extended this result over a wide range in halo masses.

The median MAH from the full MW sample is plotted as the dashed line in
Fig.~\ref{fig:mah}; the gray shaded region marks the $\pm 1\,\sigma$ region for
the MAHs.  A purely exponential form is a {\it poor} fit to the median MAH, while the
modified form of \citeauthor{tasitsiomi2004} with $\kappa=0.795$ and
$\eta=0.435$ provides a fit to within 10\% for $z < 5$.  
The median MAH can be fitted extremely well by
\begin{equation}
  \label{eq:4}
  M(z)=M_0 \,(1+z)^{\eta}\,\exp[{-\kappa^{\prime}\,(\sqrt{(1+z)}-1)}]
\end{equation}
with $\kappa^{\prime}=4.90$ and $\eta=2.23$: this fit matches the measured
median MAH to within 1.6\% for $z \le 10$.  The solid lines in
Fig.~\ref{fig:mah} show the MAHs for the Aquarius halos measured from the
\millen.\footnote{Figure 13 of \citet{boylan-kolchin2009} shows that there is
  excellent agreement between the MAHs of the level 2 Aquarius simulations and
  the corresponding halos in the \millen.}  The Aquarius halos sample the full
distribution of MAHs: halos A and C form early, halo F forms late, and halos D
and E track the median MAH.  Halo B trails the mean MAH at early times but
catches up at $z\approx 2$.

While the MAHs are normalized to unity at $z=0$, the $1\,\sigma$ scatter in
$M(z_0)/M(z=0)$ grows fairly quickly with redshift, reaching a factor of 2.5 at
$z=2$ and 4 at $z=4$.  The scatter in $\log(1+z)$ at fixed $M/M(z=0)$ is nearly
constant for $M/M(z=0) \la 0.6$.  As a result, formation times $z_f$ defined
with respect to fixed fractions of the final mass -- i.e., $M(z_f)=f \,M(z=0)$
-- will all have similar scatter in $\log(1+z_f)$, approximately 0.25, for $f
< 0.6$.

While characterizing the growth of $\mvir$ for a halo that reaches MW-like
masses at $z=0$ is useful, it is far from a complete description of that halo's
assembly history.  Many studies have shown that halo growth occurs inside-out,
with the central regions built up first and the outskirts assembled later
(e.g., \citealt{fukushige2001, loeb2003, zhao2003, diemand2007a, cuesta2008}).
Furthermore, defining halo boundaries with respect to $\Delta \,\rho_c(z)$
means that a halo's virial mass will increase with time, even in the absence of
any physical accretion, simply due to the re-definition of the boundary.  A
halo can therefore grow substantially in $\mvir$ without any appreciable change
in the mass at small radii, where the main baryonic component should lie.
Accordingly, it is useful to find a way to characterize the build-up of the
central regions of halos in addition to studying the growth of $\mvir$.

To this end, we consider the growth of $\vmax^2$ in the bottom panel of
Fig.~\ref{fig:mah}.  For the sample of Milky Way-mass halos considered here,
$\rmax$ is on average a factor of 6.1 smaller than $\rvir$, so $\vmax^2$ probes
the mass distribution on a scales $\sim 6$ times smaller than $\rvir$.
Additionally, the gravitational potential energy per unit mass of a halo is
proportional to $\vmax^2$, so studying the evolution in $\vmax^2$ probes the
growth of the dark matter halo's central potential.  The dashed curve in the
bottom panel of Fig.~\ref{fig:mah} show the median relation while the shaded
region shows the $\pm 1 \,\sigma$ scatter.  For redshifts $z \le 10$, the
median value of $\vmax^2(z)$ can be approximated to within 1.3\% as
\begin{equation}
  \label{eq:max_vmax}
  \vmax^2(z)=V_{\rm max,0}^2 \,(1+z)^{0.338}\,e^{-0.301\,z}\,.
\end{equation}

Qualitatively, the trend in $\vmax^2(z)$ is the same as in $\mvir(z)$: both
grow rapidly at early times and slowly at late times.  At a more detailed
level, however, the two show important differences in growth.  $\vmax^2$ -- and
consequently, the central potential -- approaches its redshift zero value much
earlier than $\mvir$: if we define a formation redshift $z_{f,v}$ as the time
when $\vmax^2$ reaches half of its present value, then the median $z_{f, v}$ is
approximately 4, in comparison with $z_f \approx 1.2$.  This is a reminder that
the central potential of a halo is set much earlier than its virial mass.  The
scatter in $\log(1+z)$ at fixed $\vmax^2$ is noticeably larger than at fixed
$\mvir$, showing that the spread in formation times is larger if these are
defined using $\vmax^2$ rather than $\mvir$.

As in the upper panel of Fig.~\ref{fig:mah}, the build-up of $\vmax^2$ for the
Aquarius halos (the colored lines) spans the range given by the full sample of
MW-mass halos.  Furthermore, the behavior of mass growth for individual halos
is matched by that of $\vmax^2$.  Halos A and C also assemble their central
potentials at an early time, while F forms much later.  

A stable central potential is likely to be conducive to disk galaxy formation
and evolution, as potential fluctuations may drive ``secular'' evolution and
help transform disks into bulges.  Halos A-E have nearly constant central
potentials from $z=1$ to $z=0$ (in the case of C, from $z=2$), indicating they are
possible disk galaxy hosts.  Halo F, which experiences a late major merger,
would likely host a spheroid-dominated galaxy.  These expectations are
generally borne out by SPH simulations of the Aquarius halos by
\citet{scannapieco2009}.  Using a dark matter mass resolution 2 to 3 times better
than that of the \millen, they find that halos C, D, and E all have
well-defined disks and that halos A and B also contain (smaller) disks, while
halo F has a spheroid with no disk at all.

\subsection{Halo Concentrations}
\label{subsec:conc}
\begin{figure}
 \centering
 \includegraphics[scale=0.57, viewport=0 0 432 397, clip]{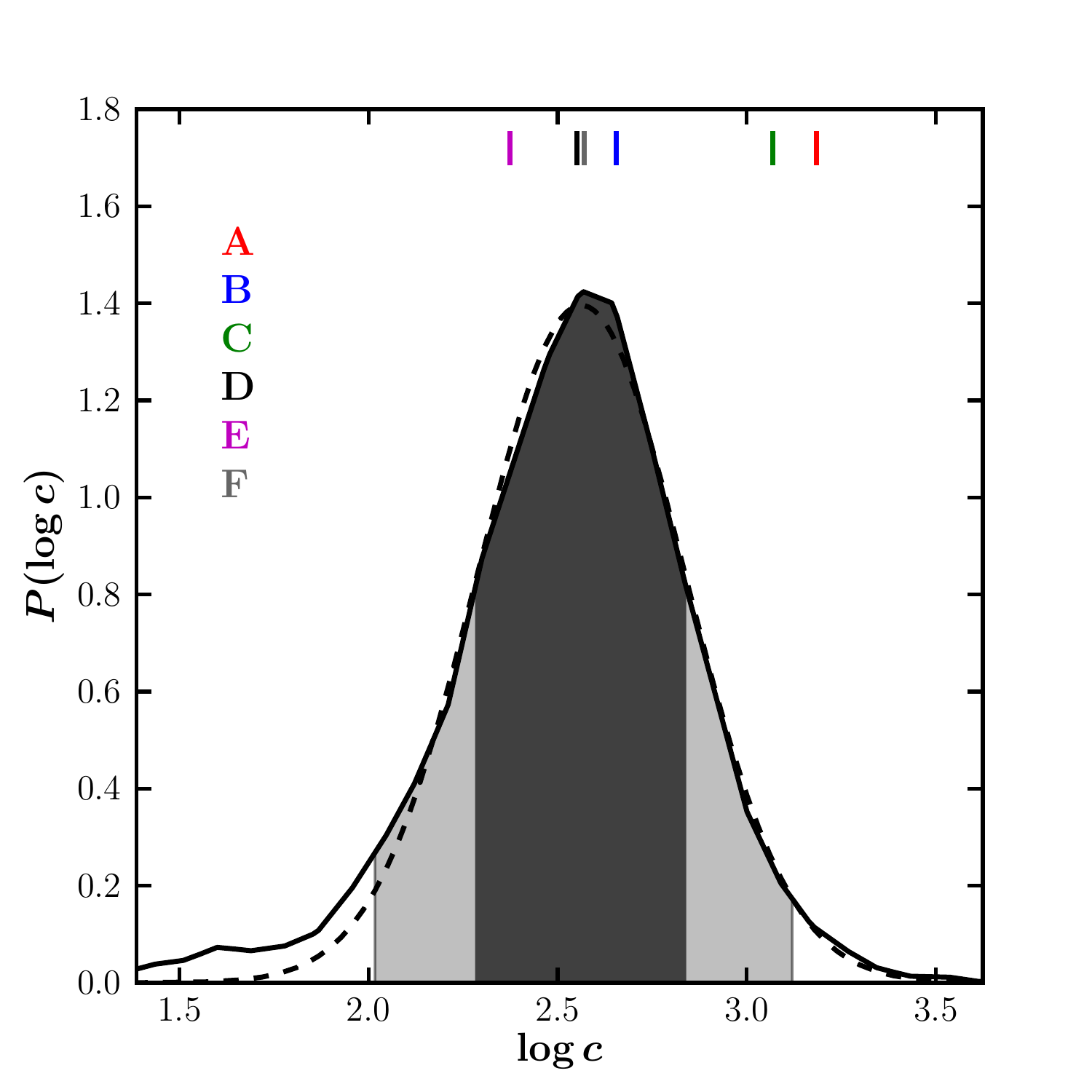}
 \caption{The probability distribution of halo concentrations $c$
  (solid black line), along with the best log-normal fit
  (dashed line) with mean $\langle \log c \rangle = 2.56$ and standard
  deviation $\sigma_{\log c}=0.272$.  Values for the individual Aquarius halos
  in the \millen\ are shown as colored vertical lines.  See the text for a
  discussion of how the concentrations are computed.
   \label{fig:conc}
 }
\end{figure}
While numerical simulations have shown that dark matter halos have a nearly
universal density profile \citep[hereafter NFW]{navarro1996, navarro1997}, the
radial scale of the profile does not follow the virial scaling of $R \propto
\mvir^{1/3}$.  An additional parameter is therefore required to specify the profile
of the typical halo of a given $\mvir$; a common choice is halo concentration $c$,
defined as the ratio of the virial radius to the radius $r_{-2}$ at which the
logarithmic slope of the density profile reaches $-2$.  The scatter in $c$ at
fixed halo mass is related to the diversity of halo formation histories.  Previous
studies have shown that average halo concentrations decrease weakly with halo
mass -- $c \propto M^{-0.1}$ -- with a fairly large scatter at fixed mass,
$\sigma(\log_{10} M) \approx 0.1-0.14$ \citep{navarro1997, jing2000,
  bullock2001, neto2007, gao2008, maccio2008, zhao2009}.

We can estimate a concentration parameter for each halo in our MW-mass sample
by assuming that each halo's density structure can be fitted with an NFW
profile; we then have $\rmax=2.16 \,r_{-2}$ and therefore, $c=2.16
\,\rvir/\rmax$.  The measured $\rmax$ values for the Aquarius halos in the
\millen\ differ from those in the level 2 Aquarius simulations by less than
10\%, showing this radius to be accurately determined at \millen\ resolution.
The probability distribution function for $\log c$ from our MW-mass halo sample
is plotted as a solid black line in Fig.~\ref{fig:conc}.  The dashed line shows
the best-fitting normal distribution with $\langle \log c\rangle=2.56$ and
$\sigma_{\log c}=0.272$.  The dark (light) shaded region marks the $\pm 1
\,\sigma$ ($ \pm 2\,\sigma$) range of this distribution.  There is a slight but
noticeable excess of halos at low concentrations compared to the best-fitting
log-normal distribution; \citet{neto2007} have shown that this is due to a
population of unrelaxed halos.

Vertical colored lines mark the concentrations of the Aquarius halos in the
\millen, computed as described above.  Halos B, D, E, and F lie near the peak
of the distribution while halos A and C lie in the high concentration tail
(though see the caveat about halo A below).  This behavior is expected given
the results of Fig.~\ref{fig:mah}: the concentration of a halo is (loosely)
related to the density of the universe when the halo formed.
Early-forming halos such as A and C are more concentrated because the 
universe was denser when they formed.

It is important to note that we did not fit the density profiles of these halos
to a specific parametric form such as NFW when estimating the concentrations
presented here.  Instead, $\rmax$ and $\rvir$ were measured directly from the
simulation data using {\tt SUBFIND}.  \citet{navarro2010} have made a thorough
study of the structure of the Aquarius halos at $z=0$, concluding that their
density profiles are not perfectly universal and therefore cannot be fully
described by a model with two scale parameters but no shape parameter (such as
NFW).  In agreement with \citet{neto2007} and \citet{gao2008}, they found
significantly better fits with the three parameter model of
\citet{einasto1965}, although the mass range spanned by the Aquarius halos is
too small to see the systematic redshift dependence of the shape parameter that
was measured in the earlier work.  We emphasize that profile fitting typically
attempts to represent the widest possible range in radius but tends to focus on
the inner regions, usually starting at the minimum resolvable radius.  This is
in part because inner regions of halos are more relaxed and also more regular
than the outskirts.  For example, Navarro et al. show that the Aquarius A-1
halo is fitted extremely well by an Einasto profile over the range $0.1\,h^{-1}
\la r \la 25\,\hkpc$ but that at larger radii, its density profile rises above
this model due to substructure and recently accreted material.  The latter is a
substantial contribution to the mass and leads to an increase in the virial
radius of the halo relative to an extrapolation of the profile fitted to the
inner region.  This is not the case for any of the other halos (see fig.~3 of
Navarro et al.)  In this sense, the measured concentration of the A halo is
larger than it ``should be.''

\subsection{Halo Spin}
\label{subsec:spin}
\begin{figure}
 \centering
 \includegraphics[scale=0.56, viewport=0 0 421 405, clip]{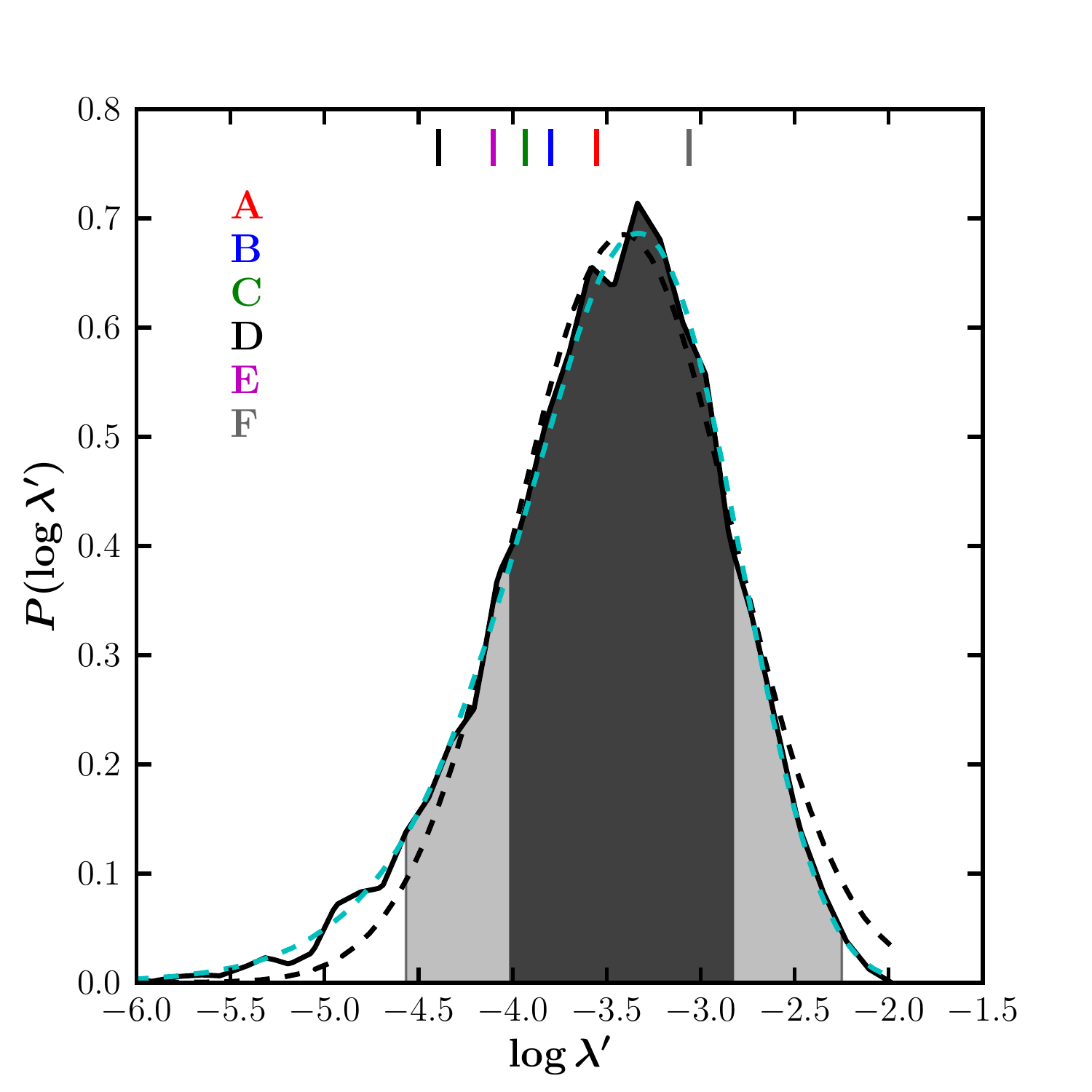}
 \caption{The probability distribution for the spin parameter
   $\lambda^{\prime}$ for Milky Way-mass halos in the \millen\ (solid black
   curve), along with $\pm 1 \, {\rm and} \, 2 \, \sigma$ regions (shaded dark and
   light gray, 
   respectively).  Dashed black line: best-fitting log-normal distribution,
   with mean $\langle \log \lambda^{\prime} \rangle = -3.41$ and standard
   deviation $\sigma_{\log \lambda^{\prime}}=0.580$.  Dashed cyan line: fit
   to the distribution given in equation~(\ref{eq:spinFit}) with 
   $\lambda_0=0.0358$ and $\chi=2.50$.
   \label{fig:spin}
 }
\end{figure}

The angular momentum content of a galaxy is often assumed to be connected to
the spin of its dark matter halo.  In the commonly-adopted model of
\citet*{mo1998}, the specific angular momentum of a galactic disk is a fixed
fraction of that of its host halo.  Many more recent versions of the Mo,
Mao, and White model (e.g., \citealt{dutton2007}) modify this assumption only
slightly.  Quantifying the angular momentum content of dark matter halos may
therefore help to connect dissipationless $N$-body simulations to the
properties of disk galaxies.
   
A convenient parametrization of a dark matter halo's angular
momentum is the modified spin parameter $\lambda^{\prime}$
\citep{bullock2001b}, defined as
\begin{equation}
  \label{eq:spin}
  \lambda^{\prime} \equiv \frac{1}{\sqrt{2}}\frac{j}{\vvir \, \rvir} \,,
\end{equation}
where $j$ is the specific angular momentum of the halo.  The distribution
of spin parameters for dark matter halos as a function of their mass has been
studied by several authors \citep{barnes1987, warren1992, cole1996, lemson1999,
  bullock2001b, bailin2005, maccio2007, bett2007, maccio2008}.  One of the most
recent, and extensive, studies is that of \citet{bett2007}, who showed that the
distribution of halo spins in the MS is well-fitted by the
function
\begin{equation}
  \label{eq:spinFit}
  P(\log \lambda^{\prime}) = A \left( \frac{\lambda^{\prime}}{\lambda_0}
  \right)^3 \,\exp \left[ -\chi \left (\frac{\lambda^{\prime}}{\lambda_0}
    \right)^{3/\chi}\right]\,. 
\end{equation}
In Figure~\ref{fig:spin}, we show the probability distribution function of
$\lambda^{\prime}$ for the \millen\ MW-mass halos (solid curve), with $\pm 1$
and $2 \, \sigma$ regions given by the dark and light shaded regions,
respectively.  The best-fitting log-normal distribution has $\langle \log
\lambda^{\prime} \rangle = -3.41$ and $\sigma_{\log \lambda^{\prime}} \!=\!
0.580$ and is shown as a dashed black line.  The dashed cyan line shows a fit
to equation~(\ref{eq:spinFit}) with $\lambda_0=0.0358$ and $\chi=2.5$; this
provides a significantly better description of the data than the log-normal
fit.

The $\lambda^{\prime}$ values for all of the Aquarius halos in the \millen\ are
shown as vertical tick marks in Fig.~\ref{fig:spin}.  While all of the halos
lie within the $2\,\sigma$ region, five of the six lie below the median
$\lambda^{\prime}$.  The lone exception, halo F, had a recent major merger (see
Fig.~\ref{fig:mah}).  Recent major merger remnants tend to have spin parameters
that are higher than typical \citep{vitvitska2002, maller2002}.  The spin
parameter of these halos usually decreases after the major merger due to
further accretion, which brings in additional non-aligned angular momentum.  It
therefore seems that at least five, and possibly all six, of the Aquarius halos
lie in the lower half of the $\lambda^{\prime}$ distribution.

\section{subhalos}
\label{sec:subhalos}
\begin{figure}
 \centering
\includegraphics[scale=0.56, viewport=0 0 421 405, clip]{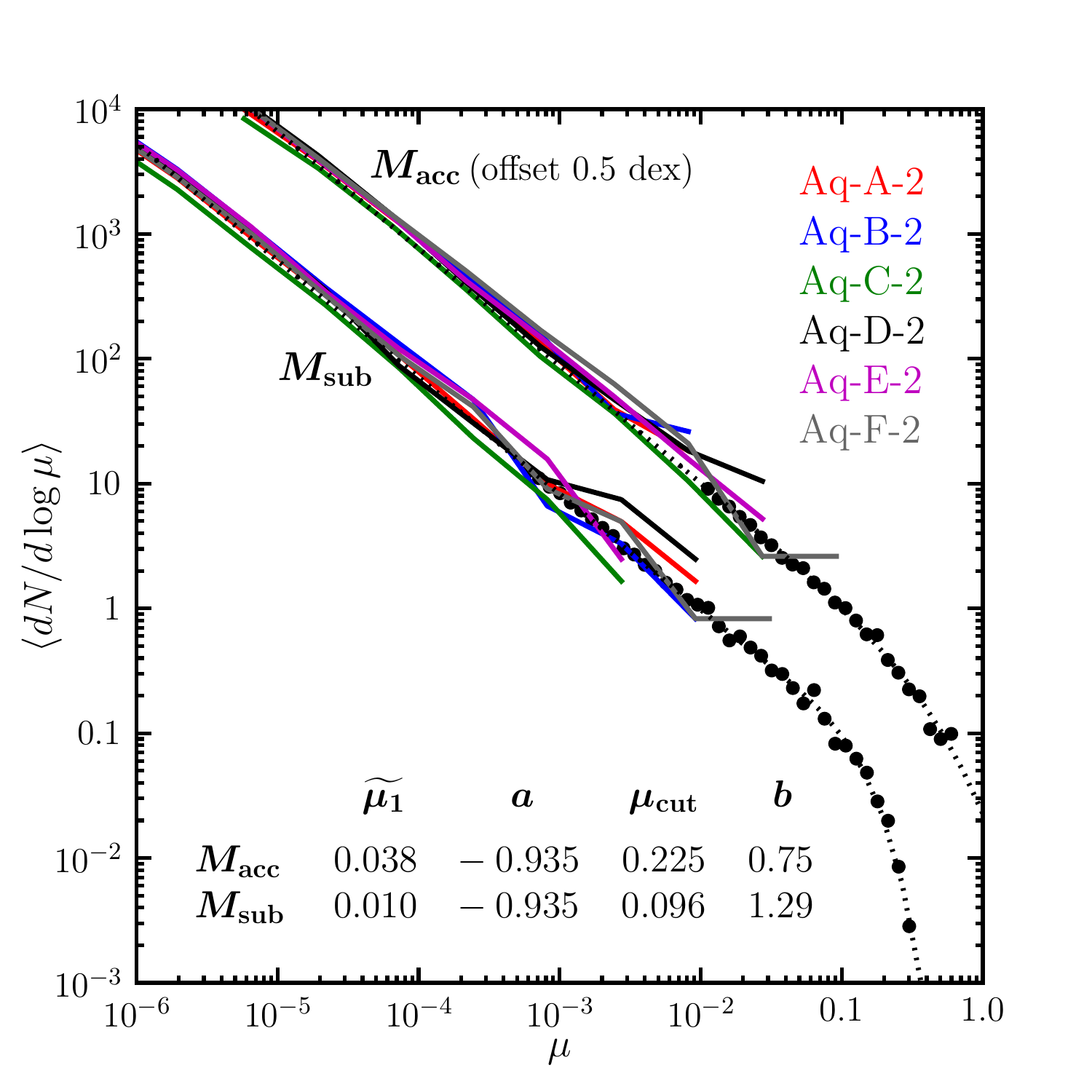}
\caption{
   The differential subhalo mass function.  Data from the 2039
   MW-mass halos in the \millen\ with $10^{12} \le \mvir / \hmsun \le
   10^{12.5}$ are shown as the black data points, while the best-fitting
   relation using equation~(\ref{eq:dmf}) is 
   shown as a dotted black line.  Results are plotted both for redshift zero
   subhalo masses ($\msub$, lower points) and for masses at accretion ($\macc$,
   upper points; offset 0.5 dex vertically for clarity).  Extrapolation of the
   \millen\ results to 
   low $\mu$ agrees well with results
   from the individual level 2 Aquarius simulations (colored lines).  
  \label{fig:dmf}
}
\end{figure}
The abundance of subhalos in MW-mass halos has been the topic of fierce debate
since late in the 1990's, when cosmological simulations were finally able to
reliably resolve dark matter substructure within galaxy-scale halos.  These
studies predicted that MW-mass halos should have hundreds of dark matter
subhalos with $\vmax \ga 10\,\kms$, while only of order ten such satellite
galaxies were seen in the Local Group \citep{klypin1999, moore1999}.  Since
then, the discovery of several faint Local Group satellites in SDSS data (e.g.,
\citealt{belokurov2007}) has extended the range of observed satellite
luminosities by almost four orders of magnitude, to $\sim 10^{3}\,L_{\odot}$,
roughly doubling the observational sample.  On the other hand, the dynamic
range of dark matter simulations has increased by an additional three decades
in mass and the number of resolved dark matter subhalos has increased by a
factor of approximately one thousand: \citet{springel2008} find $10^5$ subhalos
with $\vmax > 2 \,\kms$ inside $\rtwom$ of the Aq-A-1 simulation at $z=0$ (see
also \citealt{diemand2008} and \citealt{stadel2009}).  It is unambiguous that
the observed number of luminous satellites with $\vmax \ga 10$ falls far short
of the expected number of dark matter subhalos above the same $\vmax$ threshold
for standard \lcdm\ models.  The resolution of this discrepancy likely lies in
the strong dependence of galaxy formation physics on $\macc$ (see
\citealt{kravtsov2010} for a recent review) combined with the limited sky
coverage and the surface brightness detection threshold of the SDSS
\citep{tollerud2008, walsh2009}.  Substructure in galaxy-mass halos is
therefore still of great interest for clarifying the properties both of dark
matter and of galaxy formation at low mass (and low star formation efficiency).
In this section, we explore the subhalo content of the MW-mass halo sample from
the \millen\ at $z=0$.

In the analysis that follows, {\it the dominant subhalo is excluded in all
  cases}.  Furthermore, we consider only subhalos lying within $\rvir$; from
their host; this provides consistency when comparing halos of different host
mass.  It is very important to note that using a different limiting radius such
as $\rtwo$ or $\rtwom$ will strongly affect subhalo abundance statistics
because the radial distribution of subhalos is heavily weighted toward large
radii \citep{de-lucia2004, gao2004b, springel2008}.  This choice also affects
statistics measuring the dynamical evolution of subhalos, as subhalos falling
into their host for the first time experience tidal stripping that depends on
the local density and therefore, on the pericenters of their orbits about the
halo.

\subsection{Subhalo mass function}
\label{subsec:mass_func}
\begin{figure}
 \centering
 \includegraphics[scale=0.56, viewport=0 0 421 405, clip]{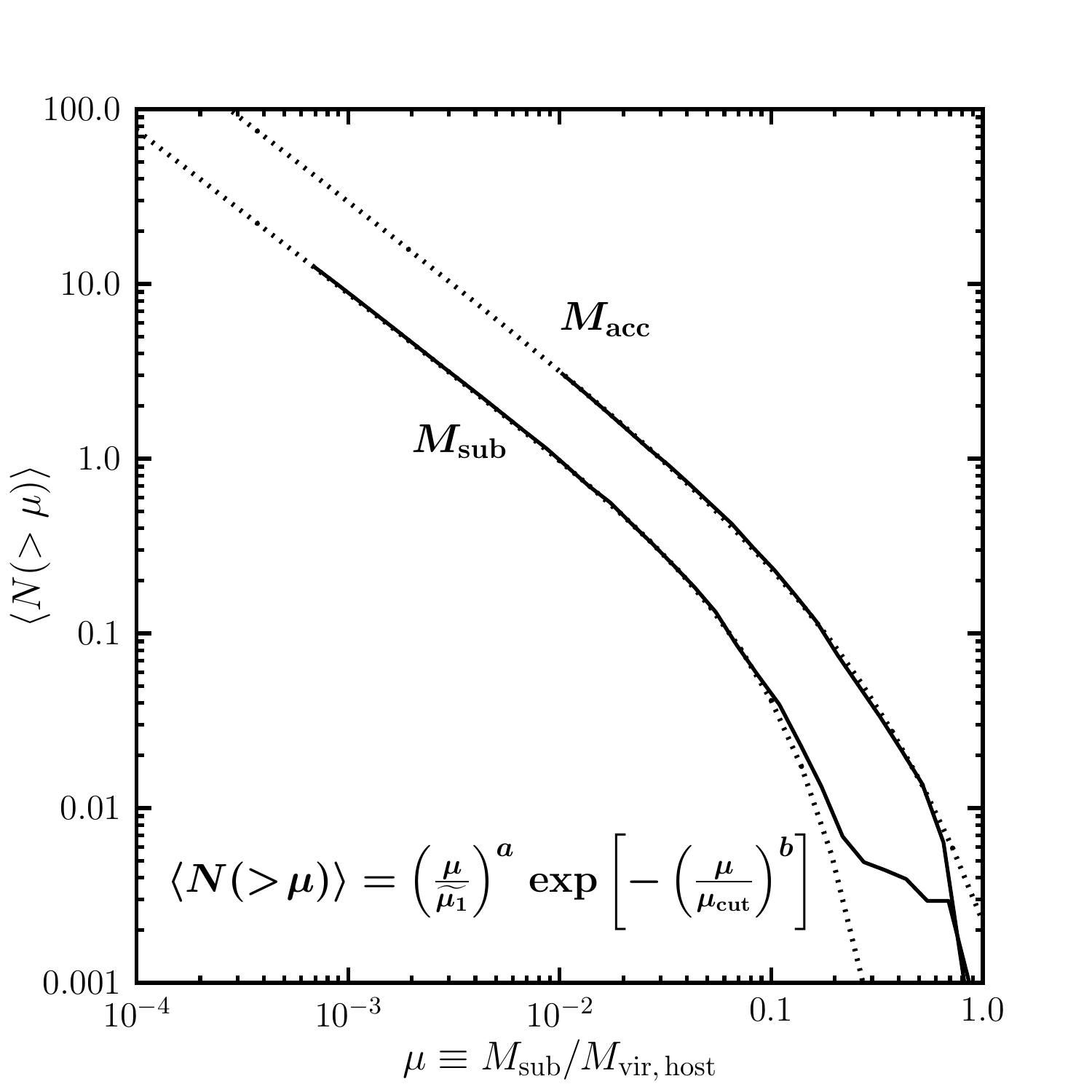}
 \caption{ The cumulative subhalo mass function for MW-mass halos having
   $10^{12} \le \mvir / \hmsun \le 10^{12.5}$.  Results are shown both in terms
   of $\msubo$ (lower solid line) and $\macc$ (upper solid line).  The dotted
   curves show equation~(\ref{eq:cmf}) using the parameters determined
   by fitting the differential mass functions (Fig.~\ref{fig:dmf}).
\label{fig:cmf}
}
\end{figure}

As shown by several authors (e.g., \citealt{gao2004b, giocoli2008,
  angulo2009}), the abundance of subhalos with a given fraction of their host's
mass varies systematically, with more massive hosts having more subhalos.  This
reflects the later formation times of more massive halos, which leave their
subhalos less time to be tidally stripped.  In the \millen, the normalization
of the differential mass function increases by approximately 15\% for each
decade in host mass.  In order to maximize our resolution, in this section we
use only the 2039 \millen\ host halos with $10^{12} \le \mvir / \hmsun \le
10^{12.5}$ when computing the mass function.

The differential mass function of (non-dominant) subhalos is shown for this
sample in Fig.~\ref{fig:dmf}. Results are plotted as black points both for the
actual redshift zero masses ($\msub$; lower data points) and also for the
accretion masses ($\macc$; upper points, offset vertically by 0.5 dex for
clarity).  For comparison, subhalo mass functions obtained from the individual
level 2 Aquarius simulations (thick colored lines) are also plotted in
Fig.~\ref{fig:dmf}.

We assume that the cumulative mass function can be represented by a
power law with an exponential cut-off at high masses:
\begin{equation}
  \label{eq:cmf}
  \cmf = \left(\frac{\mu}{\muone} \right)^a \, \exp \left[-\left(
      \frac{\mu}{\mu_{\rm cut}} \right)^b    \right] \,.
\end{equation}
The differential mass function then also has a simple functional form:
\begin{equation}
  \label{eq:dmf}
 \frac{\dd \langle N \rangle}{\dd \log \mu} =
  \left[b\left(\frac{\mu}{\mu_{\rm cut}}\right)^b -a\right]\,\cmf \,.
\end{equation}
Equation~(\ref{eq:dmf}) provides an excellent fit to the \millen\ data for
$\muone=0.01$, $\mu_{\rm cut}=0.096$, $a=-0.935$, and $b=1.29$.  This fit is
plotted as a dotted line in Fig.~\ref{fig:dmf}.  Although the range of overlap
between the \millen\ and Aquarius data is relatively small, extrapolating the
fit from the \millen\ data to low $\mu$ agrees extremely well with the Aquarius
data all the way to the Aquarius level 2 resolution limit, $\mu \approx
10^{-6}$.
\begin{figure}
 \centering
 \includegraphics[scale=0.56, viewport=0 0 421 405, clip]{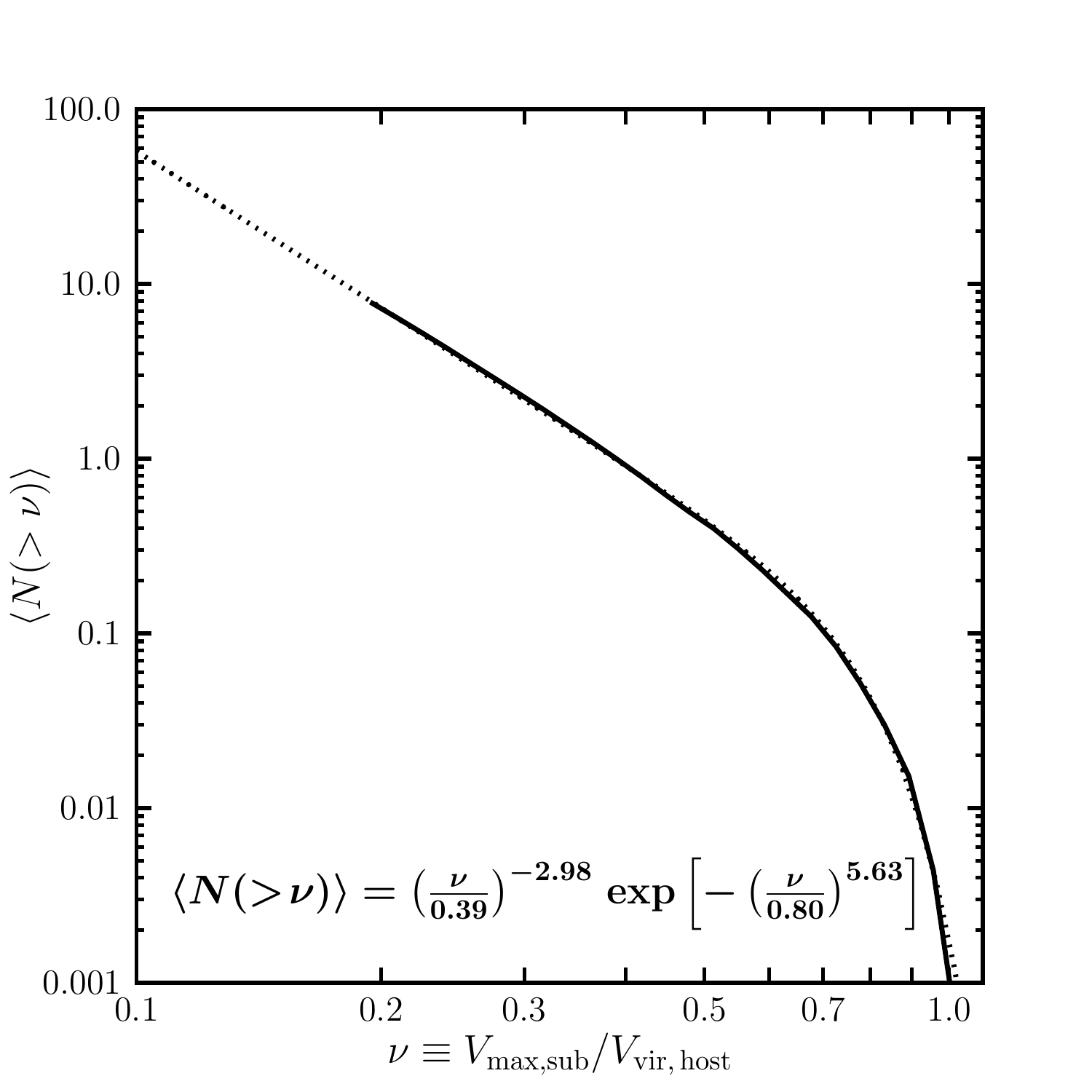}
 \caption{ The cumulative subhalo velocity function for MW-mass halos having
   $10^{12} \le \mvir / \hmsun \le 10^{12.5}$ (solid line), plotted in terms of
   $\nu \equiv \vmax/V_{\rm vir, host}$.  The dotted line shows the analog of
   equation~(\ref{eq:cmf}), determined by fitting the differential velocity
   function; the best-fitting parameters are given in the figure.
\label{fig:cvf}
}
\end{figure}

We also use equation~(\ref{eq:dmf}) to fit the subhalo mass function computed
in terms of $\macc$.  The comparison of MS and \millen\ results in
\citet{guo2010} suggests that $N_p(\zacc) \ga 1500$ per subhalo is required to
obtain converged results for $\macc$.  We are therefore able to probe $\cmfacc$
only for $\muacc \ga 10^{-2}$.  This limited range makes the determination of
the slope $a$ using the \millen\ data alone nearly impossible.
\citet{giocoli2008} have suggested that this slope is close to that of the
redshift zero cumulative mass function, and we also fit $\muacc$ with
equation~(\ref{eq:dmf}) holding $a$ fixed to -0.935, the value obtained from
fitting the differential mass function for $\msub$.  A good fit can be obtained
with $\muone=0.038$, $\mu_{\rm cut}=0.225$, and $b=0.75$ and is plotted as the
upper dotted line in Fig.~\ref{fig:dmf}.  Again, the extrapolation of the
\millen\ fit to low masses agrees very well with subhalo mass functions
computed directly from the level 2 Aquarius simulations (upper set of solid
curves).  This shows that our adopted value of $a$ is indeed appropriate.

The corresponding cumulative subhalo mass functions are plotted in
Fig.~\ref{fig:cmf} for redshift zero subhalo masses ($\msub$, lower solid
curve) and for subhalo masses at accretion ($\macc$, upper solid curve).  The
dotted lines show equation~(\ref{eq:cmf}) with parameters taken directly from
fits to the differential mass functions (i.e., we do {\it not} fit the
cumulative mass functions independently).  These curves are excellent
representations of the data for occupation numbers $\cmf \ga 0.03$.  At lower
$N$, the presence of a very small number of ongoing major mergers results in an
excess cumulative abundance compared to the fit for $\msub$.

The abundance of subhalos as a function of $\vmax$ is a related quantity, and
one that is often used when comparing simulation data to observations because
$\vmax$ is less affected than $\msub$ by the dynamical evolution of a subhalo
within its host.  Fig.~\ref{fig:cvf} shows the cumulative abundance of
subhalos, $\cvf$, in terms of $\nu \equiv \vmax/V_{\rm vir, host}$ (solid
curve).  It exhibits the same behavior as $\cmf$: a power law at low $\nu$ with
an exponential cut-off at high $\nu$.  By fitting the differential velocity
function (to ensure that errors in each bin are uncorrelated) to the same
functional form as in equation~(\ref{eq:dmf}) and converting the fit to the
cumulative velocity function, we find that the slope of $\cvf$ is -2.98 at low
$\nu$ and that $\cvf = 1$ at $\nu \approx 0.4$; this fit is shown as the dotted
line in Fig.~\ref{fig:cvf}.  This low-$\nu$ slope is in good agreement with
results from the individual, high resolution Aquarius and Via Lactea
simulations.

\subsection{Scatter in the subhalo mass function}
\label{subsec:scatter}
The halo-to-halo scatter about the mean relation in equation~(\ref{eq:cmf}) is
a matter of considerable interest for several reasons.  A number of groups have
invested considerable computational resources into simulating individual
MW-mass halos at extremely high resolution.  State-of-the-art simulations
currently use approximately one billion particles to resolve the mass
distribution within $\mvir$ \citep{diemand2008, springel2008, stadel2009}.
Such calculations are extremely expensive, however, so only a small number can
be performed.  In order to interpret their results, we must know how much
scatter is expected among halos of the same mass.  This scatter is also a
fundamental input parameter for halo occupation distribution (HOD) models
(e.g., \citealt{ma2000, peacock2000, seljak2000, scoccimarro2001,
  berlind2002}). 

\begin{figure}
 \centering
 \includegraphics[scale=0.56, viewport=0 0 421 405, clip]{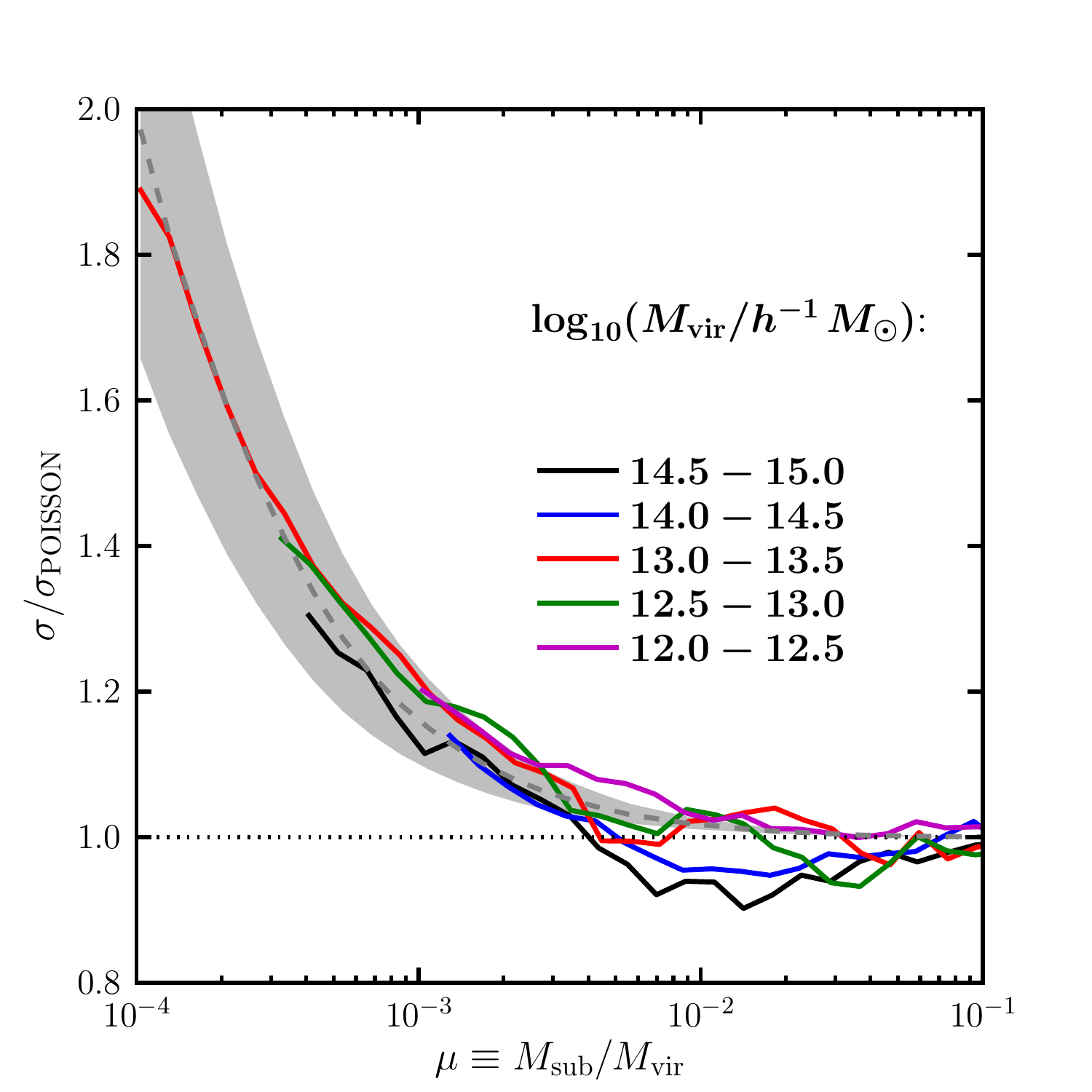}
 \caption{The ratio of the measured dispersion $\sigma$ in $N(>\mu)$ relative
   to that expected from a Poisson distribution with mean value equal to
   $\cmf$.  The dispersion is larger than Poissonian and the ratio of the two
   increases systematically 
   to lower $\mu$.  The dashed gray line shows a model in which
   the variance is a linear sum of Poisson fluctuations and an 18\% fractional
   intrinsic scatter [equation~(\ref{eq:scatter})], while the shaded gray region
   encompasses the same model with intrinsic scatter between 14 and 22\%.
  \label{fig:scatter}
}
\end{figure}
We find that the scatter in the cumulative mass function at fixed $\mu$ can be
well-modeled by postulating that the variance in $N(> \mu)$ is the sum of two
contributions, one due to Poisson fluctuations ($\sigma_{\rm P}^2 = \langle N
\rangle$) and one intrinsic ($\sigma_{\rm I}^2 \propto \langle N \rangle^2$):
\begin{equation}
  \label{eq:scatter}
  \sigma^2=\sigma_{\rm P}^2 + \sigma_{\rm I}^2 \,.
\end{equation}
This is demonstrated in Fig.~\ref{fig:scatter}.  The solid curves show
$\sigma/\sigma_P$ as a function of $\mu$ for host halos of different masses,
from Milky Way-mass hosts to halos corresponding to rich galaxy clusters.  We
have used the MS for host halos with $\mvir \ge 10^{14}\,\hmsun$, as the volume
of the \millen\ does not provide sufficient statistics at this mass scale; for all other
masses we use the \millen.  The dashed gray line shows the model of
equation~(\ref{eq:scatter}) with a fractional intrinsic scatter $s_{\rm I} \equiv
\sigma_{\rm I}/\langle N \rangle$ of 18\%, while the shaded gray region
corresponds to varying $s_{\rm I}$ from 14\% to 22\%.

This analysis shows that the scatter in the subhalo mass function is nearly
Poissonian for massive subhalos ($\mu \ga 5 \times 10^{-3}$, corresponding to
occupation numbers of $\langle N \rangle \la 2$) but is broader than Poissonian
at low masses ($\mu \la 10^{-3}$ or $\langle N \rangle \ga 8$).  This behavior
agrees with the mass functions computed from the level 2 Aquarius simulations,
which probe an additional three decades in mass to $\mu \approx 10^{-6.5}$.
The scatter in the subhalo abundance among these simulations is fairly small:
the fractional scatter, defined as the standard deviation in $N(>\mu)$
normalized by the mean of $N(>\mu)$, is constant over the range $10^{-6} \la
\mu \la 10^{-4.5}$ at approximately 12\%.  This is somewhat smaller than our
estimate from \millen\ data but is consistent given the small number of
Aquarius halos.  Poisson scatter would give 4.5\% at $\mu=10^{-5}$ and 1.5\% at
$\mu=10^{-6}$.  This is clearly inconsistent with the Aquarius results.

Many previous studies have analyzed subhalo occupation statistics using reduced
moments $\{\alpha_j \}$ of the subhalo mass function, defined via
\begin{equation}
  \label{eq:alpha}
  \alpha_j(\mu) \equiv \frac{\langle N \,(N-1) \dots (N-j+1) \rangle^{1/j}} {\langle
    N \rangle} \;\; (j =2, 3, 4, \dots)\,.
\end{equation}
These studies, which have usually been done in the context of HOD modeling,
have used the $\alpha_j$ values as measures of how close the scatter in subhalo
count is to Poissonian: for a Poisson distribution, $\alpha_j=1$ for all $j$.
All studies have found that $\alpha_2 \approx 1$ over a wide range in mass and
have concluded that the scatter is indeed Poissonian (e.g.,
\citealt{kravtsov2004a, zheng2005}).  While the finding that $\alpha_2 \approx
1$ does indeed justify the replacement of $\langle N \,(N-1) \rangle$ with $\langle N
\rangle^2$ when computing correlation functions in HOD modeling, it
does {\it not} show that the variation about $\langle N \rangle$ at fixed $\mu$
can be well-described by Poisson statistics: distributions that differ
substantially from Poisson can result in mere percent-level variations from
$\alpha_2=1$.

\begin{figure}
 \centering
 \includegraphics[scale=0.56, viewport=0 0 421 405, clip]{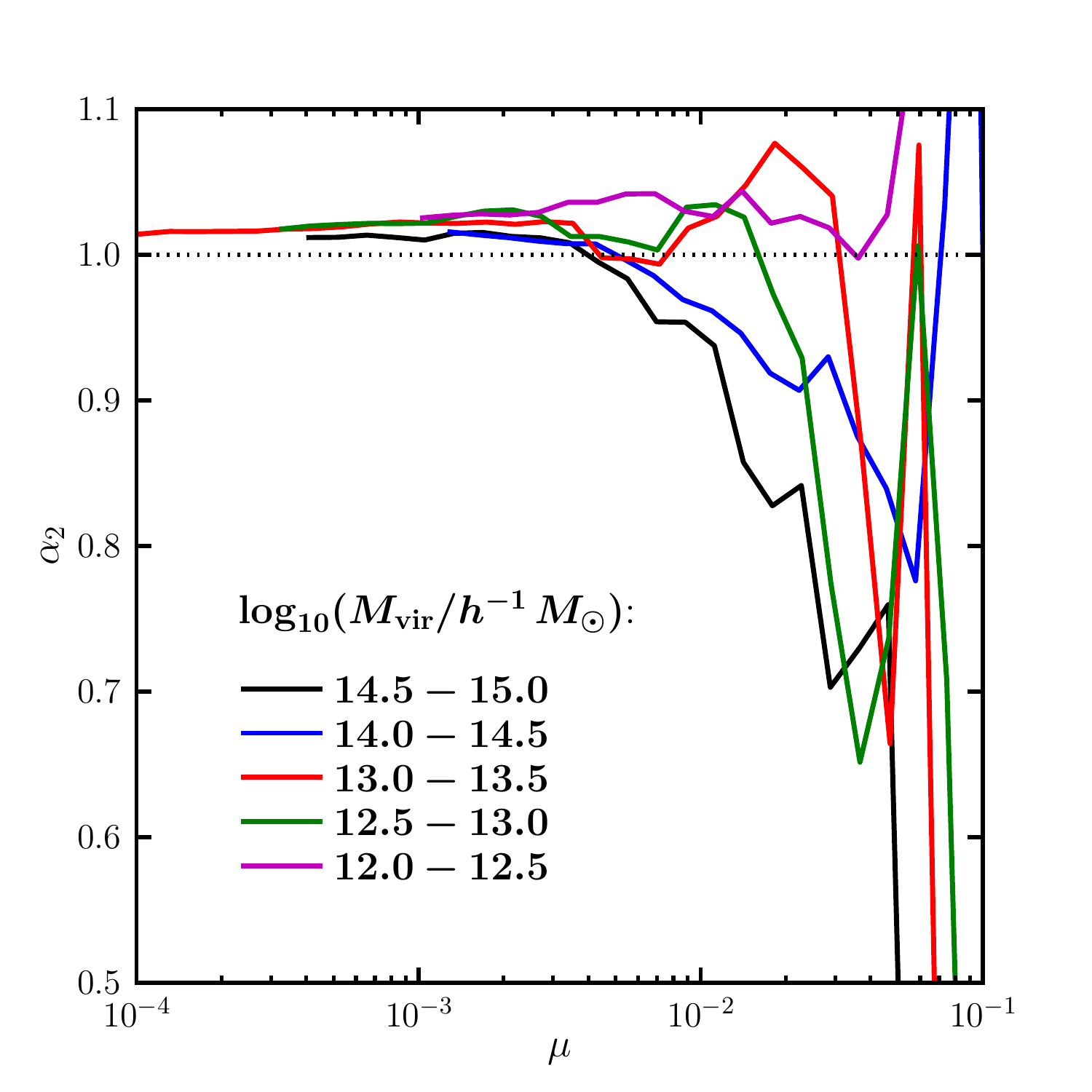}
 \caption{ The reduced second moment parameter $\alpha_2$ versus $\mu$ for
   several ranges of $\mvir$.  The behavior of $\alpha_2(\mu)$ is nearly
   independent of host mass: $\alpha_2 = 1+\epsilon$, with $\epsilon \approx
   0.02$, for $\mu \la 4 \times 10^{-3}$.  Despite this, a comparison with
   Fig.~\ref{fig:scatter} shows that the satellite HOD is {\it not} Poisson at these
   mass fractions.
   \label{fig:alpha2mass}
}
\end{figure}

Figure~\ref{fig:alpha2mass} shows $\alpha_2$ as a function of $\mu$ for the
same ranges of host halo mass as Fig.~\ref{fig:scatter}.  The behavior is
similar at all masses: $\alpha_2$ is very close to, but slightly greater than,
unity for $\mu \la 4 \times 10^{-3}$, corresponding to occupation numbers
$\langle N \rangle \ga 2$.  The deviations from $\alpha_2=1$ are quite small --
2 to 3\% -- over this range.  This is precisely the range where the scatter in
$N(>\mu)$ deviates systematically and substantially from the Poisson
expectation: Fig.~\ref{fig:scatter} shows that the scatter is 30\% broader than
Poissonian at $\mu=4 \times 10^{-4}$ and nearly twice as broad at
$\mu=10^{-4}$, whereas $\alpha_2 =1.02-1.025$ over this same mass range.  The
satellite HOD is certainly not Poisson for $\mu \la 10^{-3}$ even though
$\alpha_2$ is within 2\% of unity.  While Poisson scatter is an excellent
approximation for $\langle N \rangle \la 3$, it is increasingly inaccurate at
higher occupation number.

Why do we find a dispersion that is significantly broader than Poisson even
when $\alpha_2$ is only 2\% larger than unity?  Let us re-write $\alpha_2$ as
\begin{equation}
  \label{eq:alpha2}
  \alpha_2^2=1+\frac{\sigma^2}{\langle N \rangle^2}-\frac{1}{\langle N
    \rangle}=1+\frac{1}{\langle N \rangle} \left(\frac{\sigma^2}{\sigma_{\rm P}^2}-1
  \right)\,. 
\end{equation}
From this expression, it is clear that any distribution where $\sigma$
increases more slowly than $\langle N \rangle$ will lead to $\alpha_2 \approx
1$ for large values of $\langle N \rangle$.  Equation~(\ref{eq:alpha2}) also shows
that the deviation from $\alpha_2=1$ can be directly related to the fractional
scatter $s_{\rm I}$ in the model of equation~(\ref{eq:scatter}):
\begin{equation}
  \label{eq:scatter_frac}
  s_{\rm I} = \frac{\sigma_{\rm I}}{\langle N \rangle}
  =\sqrt{\alpha_2^2-1} 
\end{equation}
Even slight departures from $\alpha_2=1$ are therefore important: for example,
if $\alpha_2=1.1$, then $s_{\rm I}=0.46$, i.e., a 46\% intrinsic scatter.
These results suggest that more care is needed in treating the scatter in HOD
modeling.  It is still true that ${\alpha_j}^j$ specifies the error incurred by assuming
Poisson statistics when computing the $j^{\rm th}$ moment of the HOD, however.
Using Poisson statistics will therefore result in $\sim 5\%$ error in the
second moment and $\sim 8\%$ error in the third moment of the HOD.

\begin{figure}
 \centering
 \includegraphics[scale=0.68, viewport=0 0 369 770, clip]{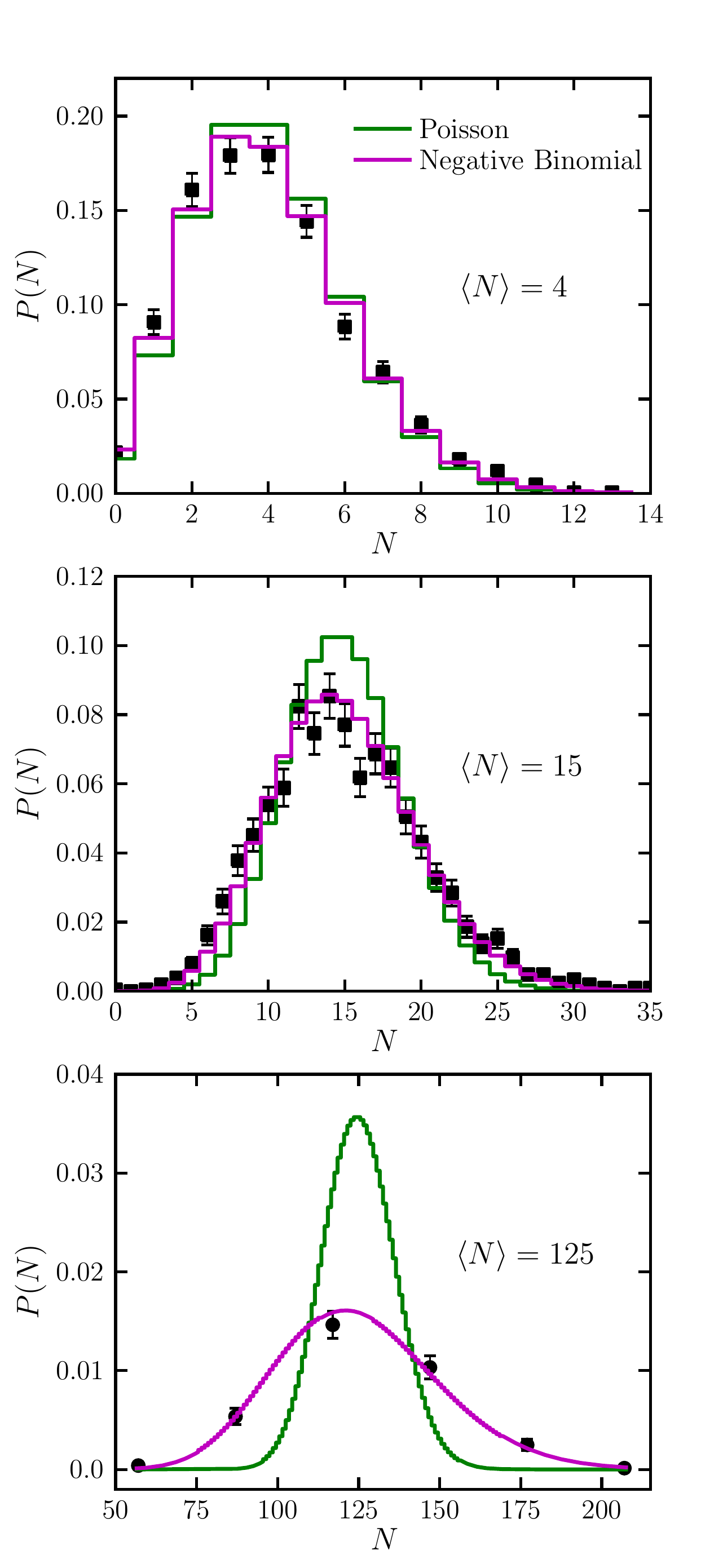}
 \caption{ The subhalo occupation distribution from host halos in the \millen\
   with $10^{12} \le \mvir \le 10^{12.5}\,\hmsun$ (upper two panels) and with
   $10^{13} \le \mvir \le 10^{13.5}\,\hmsun$ (bottom panel).  Black data points
   show the measured distribution, green lines show the Poisson distribution
   with the same $\langle N \rangle$, and magenta lines show the Negative
   Binomial distribution with the same $\langle N \rangle$ and variance given
   by equation~(\ref{eq:scatter}) with fractional intrinsic scatter $s_{\rm
     I}=0.18$.  Both distributions match the data for $\langle N \rangle=4$ (top
   panel, corresponding to $\mu > 2.3 \times 10^{-3}$), while the Poisson
   distribution is noticeably too narrow for $\langle N \rangle=15$ (middle
   panel, corresponding to $\mu > 5.6 \times 10^{-4}$) and much too narrow
   for $\langle N \rangle=125$ (bottom panel, corresponding to $\mu
   > 7.1 \times 10^{-5}$).  The Negative Binomial distribution matches the data
   well at all $\langle N \rangle$.
\label{fig:hod}
}
\end{figure}
The subhalo occupation distribution can be well-approximated by the Negative
Binomial distribution, which is given by
\begin{equation}
  \label{eq:neg_binom}
  P(N|\,r,\,p) = \frac{\Gamma(N+r)}{\Gamma(r)\,\Gamma(N+1)} \,p^r
  \,(1-p)^{N} \,.
\end{equation}
Here $\Gamma(x)=(x-1)!$ is the usual Gamma function and the parameters $r$ and
$p$ are determined by the mean $\langle N \rangle$ and variance $\sigma^2$ of
the distribution:
\begin{equation}
  \label{eq:2}
  p=\frac{\langle N \rangle}{\sigma^2}\,, \;\;r=\frac{\langle N
    \rangle^2}{\sigma^2-\langle N \rangle}  \, .
\end{equation}
Using our model for the variance in $N$, the values of $p$ and $r$ can be
expressed in terms of $\langle N \rangle$ and the fractional intrinsic scatter
$s_{\rm I}$ alone:
\begin{equation}
  p = \frac{1}{1+s_{\rm I}^2 \,\langle N \rangle}\,, \;\;r = \frac{1}{s_{\rm I}^{2}}.
\end{equation}

Given $\langle N \rangle$, it is therefore straightforward to compute the full
distribution $P(N|\,\langle N \rangle, \,s_{\rm I}^2)$.  Note that in the limit
$r \rightarrow \infty$ -- i.e., as the intrinsic scatter goes to zero -- the
Negative Binomial distribution approaches the Poisson distribution with mean
$\langle N \rangle$, as desired.

The good agreement between the measured subhalo occupation distribution and the
Negative Binomial distribution is illustrated in Fig.~\ref{fig:hod}.  Black
data points show the subhalo occupation distribution for $\cmf=4$ (top panel,
corresponding to $\mu > 2.3 \times 10^{-3}$), 15 (middle panel, corresponding
to $\mu > 5.6 \times 10^{-4}$), and 125 (bottom panel, corresponding to $\mu >
7.1 \times 10^{-5}$).  The upper two panels use \millen\ hosts with $10^{12}
\le \mvir \le 10^{12.5} \,\hmsun$, while the lower panel uses \millen\ hosts
that are a factor of ten more massive.  Poisson (green lines) provides a fairly
good match for $\langle N \rangle=4$ (the dispersion is 9\% smaller than that
of the data) but is noticeably too narrow at $\langle N \rangle=15$ (the
Poisson dispersion is 25\% smaller than that of the data).  For $\langle N
\rangle =125$, Poisson is a very poor match to the data, with a dispersion that
is too small by a factor of 2.25.  The Negative Binomial distribution (magenta
curves) matches the data well for all three values of $\langle N \rangle$.

While the results presented in this section were derived using the cumulative
mass function, we have confirmed that they hold equally well for the
cumulative maximum circular velocity function.  Specifically, the
intrinsic scatter in $\langle N(>\vmax/\vvir) \rangle$ is nearly identical to
that in $\cmf$, i.e., the subhalo abundance also shows an intrinsic scatter of
18\% when expressed in terms of $\vmax/\vvir$.  This is {\it not} true when
considering $\vmax/V_{\rm max, host}$, however.  This definition introduces
additional scatter in the subhalo abundance due to the scatter in
concentrations at fixed mass: host halos of a given $\vvir$ have a range of
$\vmax$ values.

\citet{ishiyama2009} have recently studied a sample of 125 well-resolved halos
from a cosmological simulation; their work corroborates this result.  They show
that the scatter in $N(>\!\vmax/V_{\rm max, host})$ is of order $40\%$ for host
halos with $10^{12}\la \mvir \la 2 \times 10^{12}\,\hmsun$.  They find the
scatter to be markedly reduced when using a criterion more similar to ours,
$N(>\!\vmax/V_{\rm 200m, host})$ for subhalos within $\rtwom$, signifying that
subhalo properties correlate more tightly to host $\vvir$ than to host $\vmax$
(see also \citealt{springel2008}).  This is yet another reminder that results
on subhalo mass and circular velocity functions are sensitive both to how the
host halo sample is defined and to how subhalos are selected (e.g., the
limiting radius chosen).

Finally, we have investigated whether the subhalo mass function shows
systematic variation with the host's environment (as measured by the
overdensity on a scale of $5\,\hmpc$) or host halo properties.  No
obvious correlations exist.  Furthermore, the variation in the subhalo mass
function between hosts can be modeled as a simple normalization shift, i.e.,
the mass function within individual halos is consistent with the low-mass slope
measured from the ensemble average.  This is in good agreement with the
behavior of the mass Aquarius halos' mass functions (e.g., Fig.~\ref{fig:dmf}).

\subsection{Massive subhalos}
\begin{figure}
\centering
\includegraphics[scale=0.56, viewport=0 0 421 405, clip]{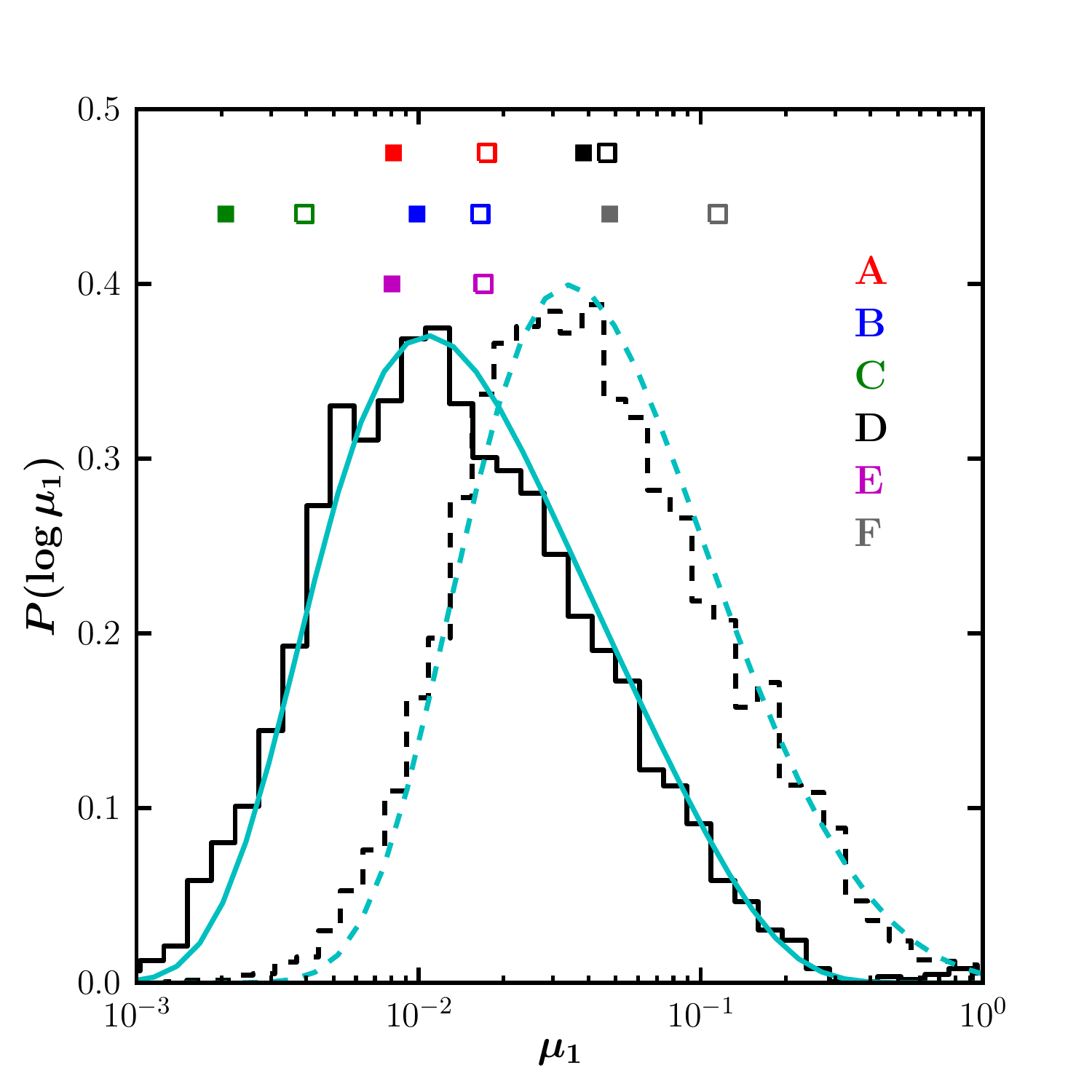}
\caption{ Distribution over $\log \mu$ of the most massive subhalo
  within $\rvir$ of each of our Milky Way-mass hosts, both in terms of
  $\msubo$ (solid histogram) and $\macc$ (dashed histogram).  In each case,
  $\mu_1$ is computed relative to $\mviro$.  The
  values of $\mu_1$ for each Aquarius halo in the \millen\ are also noted with
  square symbols (filled for $\msubo$, open for $\macc$).  The distribution of
  $\mu_1$ is fairly
  broad and peaks at $\mu_1 \approx 0.01$ (0.035) for $\msubo$ ($\macc$).  The
  cyan lines show the prediction for $\mu_1$ (solid) and $\mu_{\rm acc,1 }$
  (dashed) if the most massive subhalo is Poisson-sampled from our analytic fit to
  the cumulative
  mass function.  In both cases, the Poisson prediction agrees very well with
  the actual distribution.  
  \label{fig:mmsub}
}
\end{figure}

The parameter $\muone$ in equation~(\ref{eq:cmf}) is approximately the mass
fraction relative to $\mvir$ of the most massive subhalo in a ``typical'' host;
$\muone=0.01$ therefore shows that the most massive subhalo in a Milky Way-mass
halo typically has about 1\% the mass of its host.  The distribution of $\mu$
for the most massive subhalo in each of our hosts is shown in
Fig.~\ref{fig:mmsub}.  Data are plotted for both $\mu_1=\msubo/\mviro$ (solid
histogram and filled squares) and for $\mu_1=\macc/\mviro$ (dashed histogram
and open squares\footnote{Note that what is plotted as the dashed histogram in
  Fig.~\ref{fig:mmsub} is the probability distribution for the subhalo at $z=0$
  with the largest $\macc$, not $\macc$ of the most massive subhalo at
  $z=0$.}).  Values for the individual Aquarius halos are shown by colored
symbols.  Fig.~\ref{fig:mmsub} confirms that the probability distribution
function for $\mu_1$ peaks at $\mu_1 \approx 0.01$ for subhalo masses measured
at $z=0$, albeit with a large spread: the probability for $\mu_1$ to be larger
than 0.025 is 29\% and to be larger than 0.1 is 4.3\%, while 4.8\% of halos
have $\mu_1 < 0.0025$.

The model presented in Sec.~\ref{subsec:scatter} suggests the scatter in the
subhalo mass function should be nearly Poissonian for masses relevant for the
most massive subhalo.  We can test this directly: if the most massive subhalo
in each halo is Poisson sampled from the underlying $\cmf$, then the
probability density $P(\mu_1)$ for finding $\mu_1$ in the range
$[\mu_1, \mu_1+\dd \mu_1]$ can be directly computed as
\begin{equation}
  \label{eq:mmsub}
  P(\mu_1) \,\dd \mu_1 = \frac{\dd}{\dd \mu_1} \, e^{-\langle N(>\mu_1)
    \rangle} \, \dd \mu_1 \,.
\end{equation}
The cyan lines in Fig.~\ref{fig:mmsub} show $P(\mu_1)$ computed this way for
$\msubo$ (solid line) and $\macc$ (dashed line); the agreement with the actual
distributions for $\mu_1$ (the histograms) is excellent.  The distribution of
$\mu_1$ is therefore consistent with the Poisson hypothesis, both for the
redshift zero masses and for the accretion masses.

The distribution of $\mu_1$ computed with respect to $\macc$ can be used to
estimate the probability that the Milky Way should host a galaxy at least as
massive as the Large Magellanic Cloud (LMC), the most luminous MW satellite.
Using the abundance matching results of \citet{guo2010} and a stellar mass of
$2.5 \times 10^{9}\, \msun$ for the LMC \citep{kim1998}, we find that
$\macc({\rm LMC})=1.9 \times 10^{11} \, \msun$.  For a Milky Way mass of
$10^{12}\,\msun$, there is a 8\% chance of having a satellite with this
accretion mass or greater.  For a mass of $2.5 \times 10^{12} \,\msun$ --
consistent with the abundance matching results of Guo et al. and the Local
Group timing argument value obtained by \citet{li2008a}, this probability rises
to 27\%.

This same line of reasoning can be used to compute the probability of a dark
matter halo hosting two satellite galaxies at least as massive as the Small
Magellanic Cloud (SMC).  The SMC has a total stellar mass of approximately $3
\times 10^{8}\,\msun$ \citep{stanimirovic2004}, corresponding to an abundance
matching mass at accretion of $8.2 \times 10^{10} \msun$.  If the Milky Way has
a dark matter halo of mass $10^{12} \,\msun$, then there is a 3.3\% chance of
having a second-ranked satellite with at least this infall mass.  For a halo
mass of $2.5 \times 10^{12} \,\msun$, the probability becomes 20\%.  

These probabilities ignore any scatter in the $\mstar-M_{\rm halo}$ relation.
This scatter is subdominant to the uncertainty in the Milky Way's mass,
however.  For example, \citet{guo2010} give 80\% confidence intervals that
correspond to $\pm 25\%$ on the infall halo masses of the Magellanic Clouds
based on a dispersion of 0.2 in $\log_{10}(M_{\rm halo})$ at fixed
$\mstar$. Clearly, the satellite statistics quoted above give some support for
the higher estimates of the Milky Way's halo mass.

\subsection{Subhalo accretion times}
\begin{figure*}
 \centering
 \includegraphics[scale=0.65, viewport=10 0 657 297, clip]{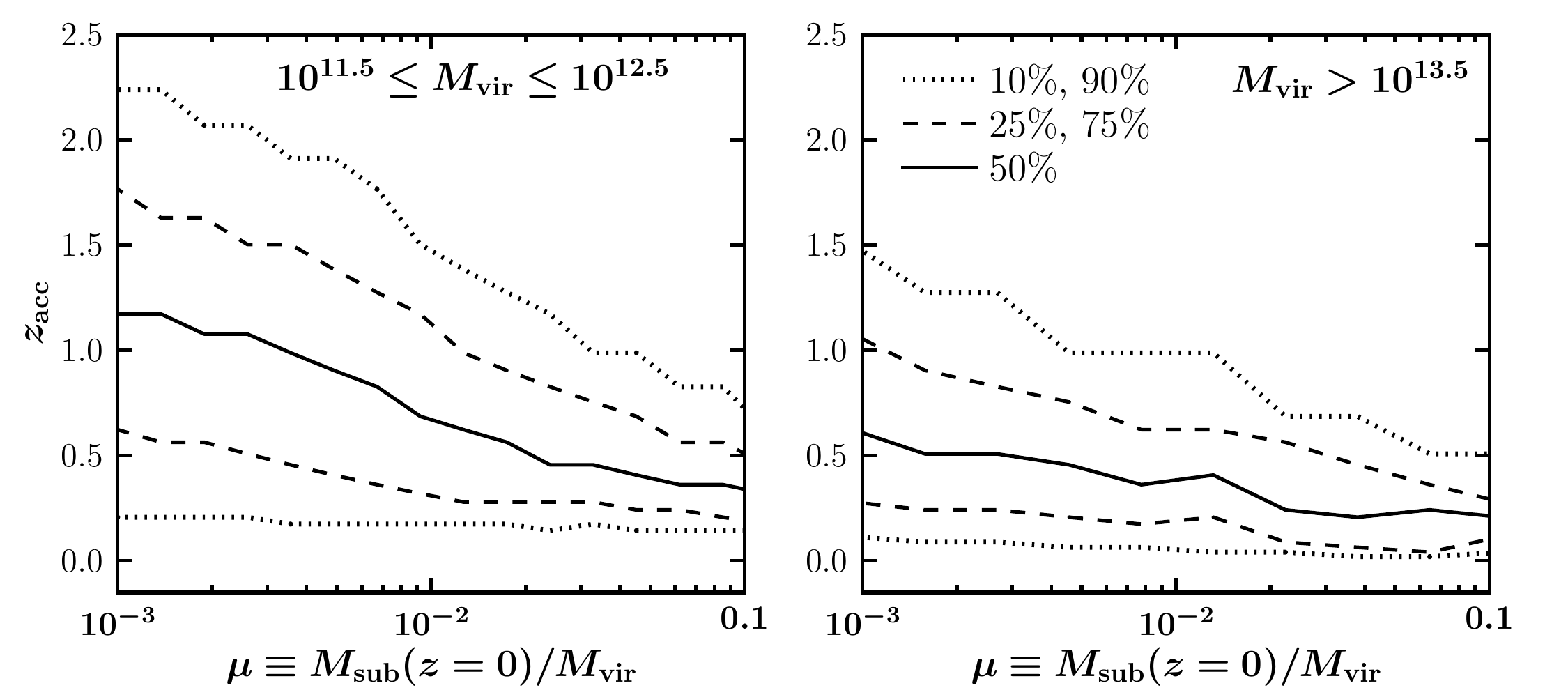}
 \caption{ Redshift of accretion $\zacc$ as a function of $\msubo/\mvir$ for
   all resolved subhalos in galaxy-mass (left) and cluster-mass (right) host halos.
  The median
   relation is shown as a solid curve while the 25 and 75 (10 and 90)
   percentiles are shown as dashed (dotted) curves.  More massive subhalos at
   the present day were accreted later on average than low-mass subhalos; the
   spread of accretion times for massive objects is also smaller.  Subhalos of more
   massive host halos are accreted at lower redshifts on average, reflecting
   the hierarchical build-up of dark matter halos.
   \label{fig:z_acc}
 }
\end{figure*}
Subhalos persisting to $z=0$ are fossils of previous halo merger events.  The
accretion redshifts $\zacc$ of $z=0$ subhalos are therefore important
quantities for understanding the dynamical evolution of subhalos and can help
constrain other quantities, such as merging timescales.  The median $\zacc$ for
$z=0$ subhalos as a function of $\mu=\msubo/\mvir$ is shown as a solid curve in
Fig.~\ref{fig:z_acc}.  The 25th and 75th (10th and 90th) percentiles of the
$\zacc$ distribution are included as dashed (dotted) curves.  Results for Milky
Way-mass host halos are shown in the left panel.  The median $\zacc$ is a
decreasing function of $\mu$, which is mainly a dynamical effect: massive
subhalos that are accreted at early times are able to lose their orbital energy
and angular momentum and merge with the dominant subhalo by $z=0$ while
low-mass subhalos are not (e.g., \citealt{boylan-kolchin2008}).  As a result,
the only remaining massive subhalos are those that were accreted at relatively
late times.  At $\mu=10^{-3}$, the typical $\zacc$ of a surviving subhalo is
somewhat larger than $z=1$, while at $\mu=10^{-1}$, $\zacc \approx 0.3$ for
surviving subhalos.  It is difficult to compare the individual Aquarius halos,
as there are only 60 satellites across all six halos satisfying $\mu \ge 0.001$
and $d \le \rvir$, but the median accretion time for these 60 satellites,
$\zacc=0.76$, falls well within the expected range based on the distribution
from the \millen.

For comparison, the right panel of Fig.~\ref{fig:z_acc} shows the distribution
of $\zacc$ for surviving subhalos in much more massive hosts, those with $\mvir
> 10^{13.5}\,\hmsun$.  The trend is very similar to that seen for Milky
Way-mass hosts, namely, the median $\zacc$ is a decreasing function of $\mu$.
The typical $\zacc$ values are uniformly lower for the massive halo sample,
however: at $\mu=10^{-3}$, the median surviving subhalo in the massive host
sample was accreted at $\zacc \approx 0.6$.  This difference is a manifestation
of the hierarchical growth of dark matter halos: more massive halos are
dynamically younger and have accreted a larger fraction of their mass (and of
their subhalo population) recently.  Fig.~\ref{fig:z_acc} also indicates
that a large fraction of the galaxy population in galaxy clusters has joined
the cluster fairly recently, whereas most of the luminous satellites of
galaxies similar to the Milky Way have been part of their host halos since $z
\approx 1$.  At all masses, the median value of $\macc/\msubo$ is 2.4 (see
Fig.~\ref{fig:z0vmax_vs_mmax}), indicating that a typical $z=0$ subhalo has
lost 60\% of its mass.

The results in the left panel of Fig.~\ref{fig:z_acc} appear to differ
substantially from those of \citet{gao2004b}, who found that 90\% of subhalos
in a Milky Way-mass halo at $z=0$ have $\zacc < 1$.  These differences are due to
the choice of definition for $\zacc$.  Gao et al. define $\zacc$ as the {\it
  last} time a subhalo entered the FOF group of the main progenitor of its
$z=0$ host while we define $\zacc$ to be the redshift when the subhalo's mass
was at its maximum.  This latter definition is closer to the {\it first} time a
subhalo enters its host's FOF group.  We have checked that using Gao et al.'s
definition of $\zacc$ gives results similar to theirs.  Our results for more
massive halos (the right panel of Fig.~\ref{fig:z_acc}) agree with the findings
of \citet[e.g., their fig.~10]{de-lucia2004}, who use the same definition of
$\zacc$ as we do.

\section{Merger histories}
\label{sec:mergers}
Galaxy-mass dark matter halos are constantly bombarded by smaller halos in the
\lcdm\ model.  This has been the source of great concern with respect to the
formation of MW-like galaxies: how is it possible to reconcile an active
merging history with the existence of a thin stellar disk (e.g.,
\citealt{toth1992, velazquez1999, benson2004, kazantzidis2008, hopkins2009})?
In order to understand the severity of the problem, we investigate the merger
histories of MW-mass halos in this section, concentrating on mergers of
objects with mass comparable to that of the MW disk.

When considering the merger history of a dark matter halo, two major questions
are (1) what is the rate of infall of other, smaller dark matter halos across
the halo's virial radius and (2) what is the rate at which other dark matter
halos lose their identity and merge with the {\it center} of the halo?  Clearly,
the set of objects that merge into the center of a halo will be a subset of
those that crossed the halo's virial radius at earlier times, but connecting
the two is far from trivial.  Several recent papers have addressed the
frequency of halo-halo mergers.  \citet*{fakhouri2008}, \citet{guo2008}, and
\citet{genel2009} investigated the halo merger rate directly in the MS, while
\citet{neistein2008} and \citet{cole2008} used extended Press-Schechter theory
to derive merger rates and compared the results to the MS.  The most relevant
study for this paper is \citet{stewart2008}, who used an $N$-body simulation to
extract halo-halo merger rates for MW-mass halos.  Stewart et al. found that
while recent major mergers are very rare, mergers involving halos with masses
of $\sim 0.1 \,\mviro$ are quite common over the past 10 Gyr and mergers with
halos having masses exceeding the mass of the galactic disk are virtually
inevitable over the same time period.  These results led the authors to
conclude that the frequency of mergers may present a substantial challenge to
our understanding of disk galaxy stability, as the majority of galaxies in
MW-mass halos are disk-dominated \citep{weinmann2006, van-den-bosch2007,
  park2007}).

The mass resolution of the \millen\ allows us to expand on the results of
Stewart et al. by studying mergers with the central regions of MW-mass halos.
These are more clearly relevant to the survival of stellar disks (which have
characteristic radii that are less than 10\% of $\rvir$).  Accordingly, {\it in
  this section we consider only mergers between accreted subhalos and the
  dominant subhalos in MW-mass halos.}  We construct samples of merging objects
by searching the main progenitor branches of the dominant subhalos for merger
events with subhalos satisfying $\vacc \ge 50 \, \kms$.

In order to ensure that the mergers we are considering are well resolved and are
not affected by limited numerical resolution, we have computed the orbits and
internal properties of subhalos immediately prior to merging.  For the mergers
considered here -- those with $\vacc > 50 \,\kms$ for the satellite -- the
median separation between the satellite and the dominant subhalo immediately
prior to merger is about $15 \,\hkpc$; in 80\% (90\%) of the cases, the
separation is less than $30 \,(50)\,\hkpc$.  

Furthermore, the satellites themselves are well-resolved immediately prior to
merging: as we show in the Appendix, even halos at our lower limit of
$\vacc=50\,\kms$ have on average 70 particles just prior to merging, while
satellites with $\vacc=65\,\kms$ typically have 150 particles at the time of
merging.  Accreted halos that are capable of substantially impacting the
Galactic disk can therefore be reliably tracked to within $\la 10\%$ of
$\rvir$, allowing us to separate true mergers with the centers of halos from
accretion events that cross $\rvir$ but remain at large halo-centric distances.

\begin{figure}
 \centering
 \includegraphics[scale=0.56, viewport=0 0 421 405, clip]{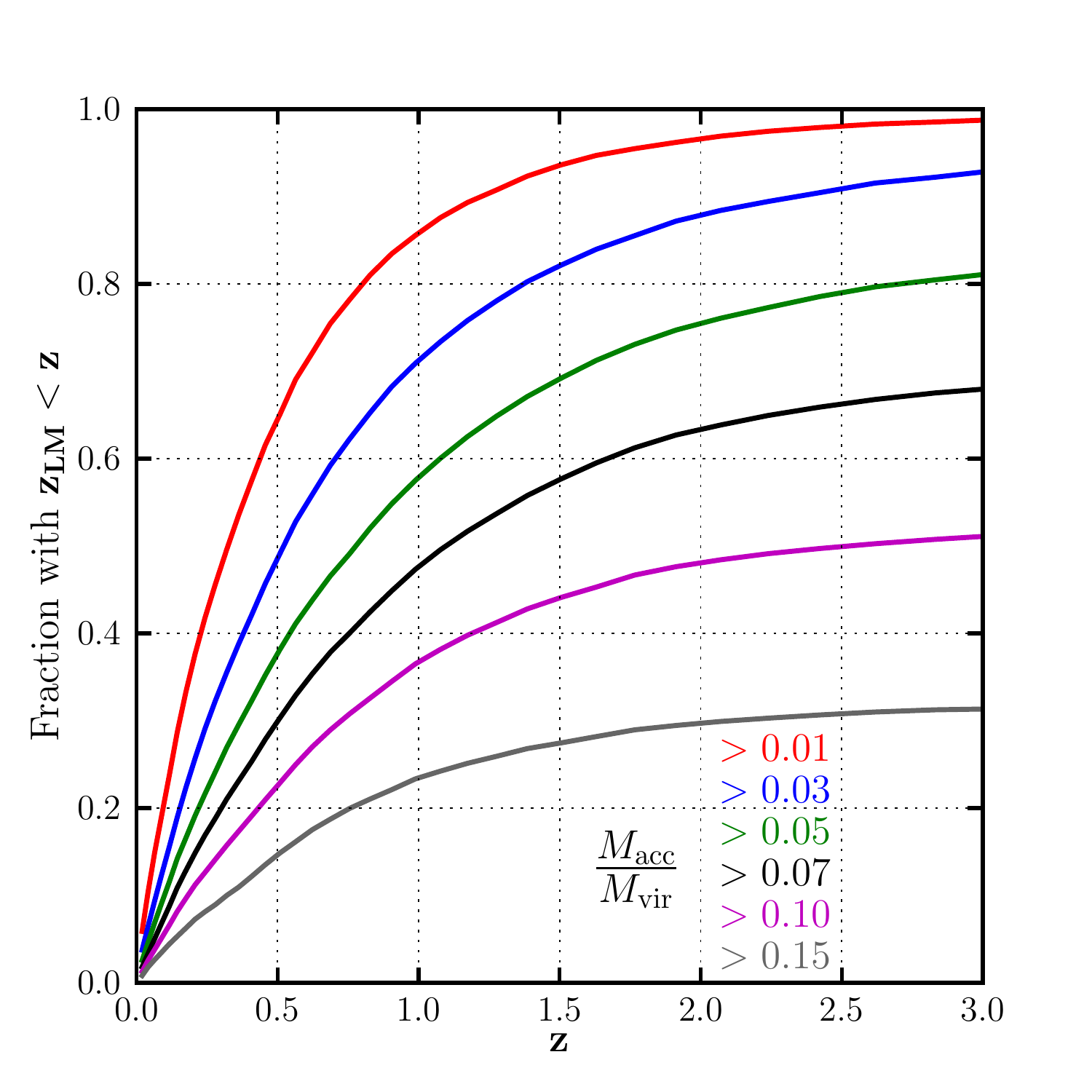}
 \caption{
   \label{fig:mergerHistory}
   The fraction of central subhalos that have merged with an object having $\macc$
   greater than 1, 3, 5, 7, 10, and 15\% of $\mviro$ (upper through lower
   curves) since redshift $z$, as a function of $z$.  Virtually all halos have
   merged with a $\macc > 0.01 \,\mviro$ halo since $z=3$, while only half have
   merged with a $\macc > 0.10 \,\mviro$ halo in the same redshift range.
   Recent merger events with $\macc > 0.03\,\mviro$ are common: $70\%$ of halos
   have had such a merger since $z=1$.  
 }
\end{figure}
Figure~\ref{fig:mergerHistory} explores the probability for dominant $z=0$
MW-mass subhalos to have merged since redshift $z$ with another subhalo with
$\macc/\mviro >0.01, 0.03, 0.05, 0.07, 0.1, \,{\rm and } \, 0.15$ (top to
bottom).  Approximately two-thirds of MW-mass halos have experienced a merger
with a halo having $\macc/\mviro > 0.03$ since $z=1$, and over 90\% have had at
least one such merger since $z=3$.  On the other hand, only 30\% of halos have
merged with a $\macc/\mviro >0.1$ halo since $z=1$ and just 50\% of halos have
ever experienced such a merger.  Approximately 50\% of halos have had a merger
with $0.03 \la \macc/\mviro \la 0.15$ since z=1.

It is important to note that we have been quoting satellite masses in terms of
the maximum dark matter mass the satellite has ever attained, $\macc$.
Subhalos usually lose a large fraction of their mass before merging: for our
sample, the average subhalo loses 90\% of its mass between accretion and
merging (see Fig.~\ref{fig:mergerMass}), independent of $\macc$.  Subhalos with
$\macc/\mviro=0.1$ therefore typically have a mass ratio of 1:100 at merger,
meaning the mass involved in a potential disk impact may be three times smaller
than the disk mass rather than three times larger.

\begin{figure}
 \centering
 \includegraphics[scale=0.56, viewport=0 0 421 405, clip]{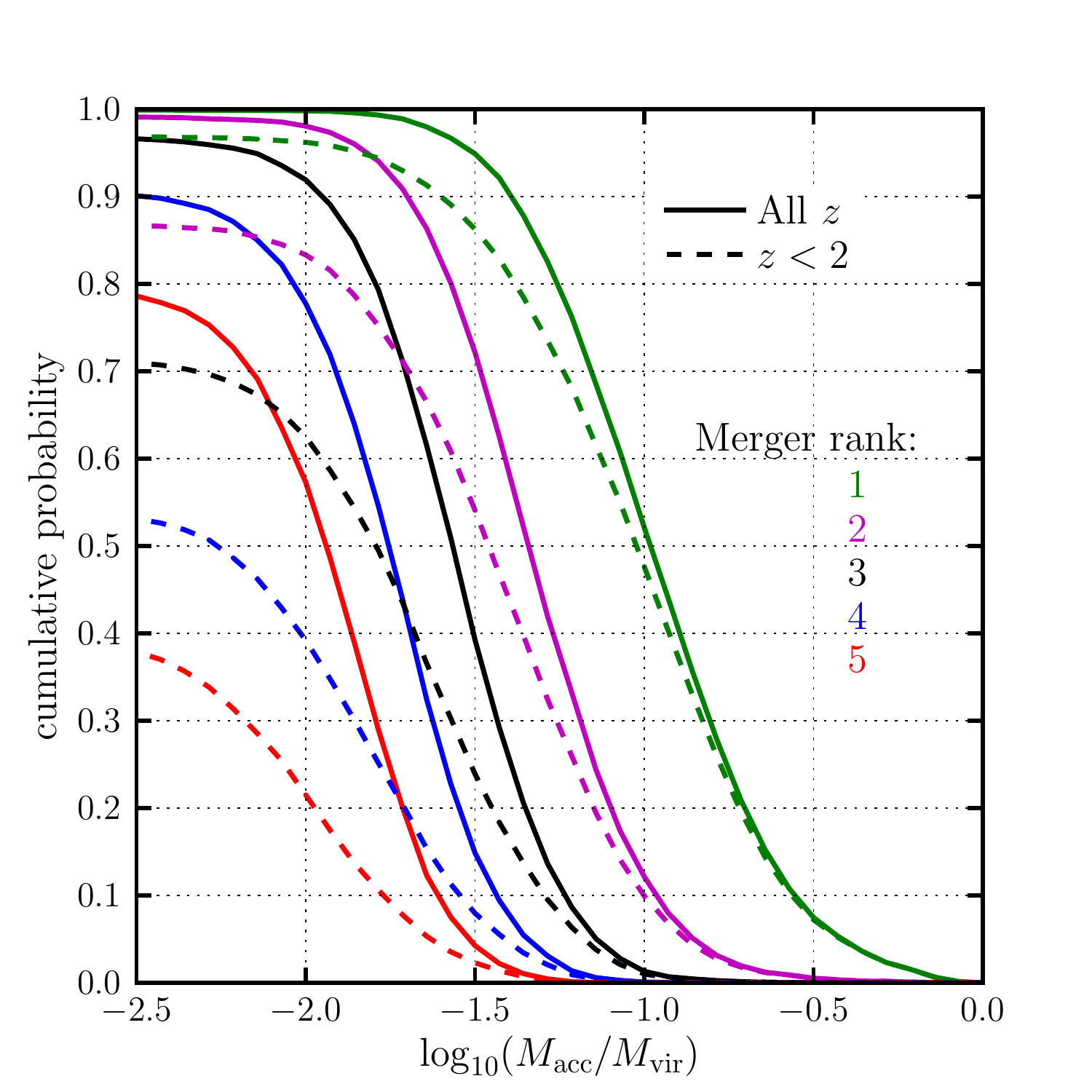}
 \caption{
   \label{fig:maccMvir}
   The cumulative probability of having a merger with $\mu > \macc/\mviro$ for
   the 1st, 2nd, 3rd, 4th, and 5th-most massive central merger that a given
   halo experiences, both over the halo's entire lifetime (solid lines) and
   since $z=2$ (dashed lines).  The median value for the most massive merger is
   $\mu=0.1$, while the 3rd-most massive merger typically has $\mu=0.02$.}
\end{figure}
We can also investigate the distributions of merger histories in terms of the
${\rm n^{th}}$-most massive halo with which a MW-mass halo has merged.  This is
done in Fig.~\ref{fig:maccMvir}.  The solid lines show the probability of
having a merger with $\macc/\mviro \ge \mu$ as a function of $\mu$, while the
dashed curves give the same probability for mergers at $z < 2$.  Different
colors correspond to different merger ranks: green shows the probability
distribution for the most massive merger for each halo (in terms of $\macc$)
while red shows the distribution for the 5th most massive merger.  90\% of
halos have had one merger event with $\macc > 0.03\,\mviro$ since $z=2$, 60\%
have had two such events, and 30\% have had three.  The median value
for the most massive merger a halo has experienced since $z=2$ is
$\macc/\mviro=0.1$, and the median redshift for such mergers\footnote{This is the
  redshift at which the satellite subhalo merges with the central subhalo, not
  the redshift at which the progenitor halos merge.} is $z=0.75$; only 10\% of
halos have had two events with $\macc/\mviro \ge 0.1$.  The most massive merger
event since $z=2$ for each of the Aquarius halos (A-F) has $\macc/\mviro$ of
0.033, 0.078, 0.028, 0.030, 0.033, and 0.259, respectively.  Halos A, C, D, and
E therefore lie near the 10\% most quiescent halos of Fig.~\ref{fig:maccMvir}
while halo B is fairly typical and halo F has had a recent major merger, which
is rare.

A thorough understanding of the stability of galactic disks in the presence of
satellite infall requires the inclusion of baryonic physics, as several studies
have shown that factors such as the orbits and gas content of the merging
galaxies can have a substantial impact on remnant properties
\citep{springel2005c, kazantzidis2008, kazantzidis2009, scannapieco2009,
  hopkins2009, purcell2009, stewart2009a, moster2010a, governato2009,
  donghia2010} Nevertheless, we can still use the statistics computed above to
learn about the frequency of satellite-disk interactions, as the dynamics that
determine which subhalos merge are mostly set by dark matter halo properties
such as the satellite's orbit and mass at infall.

Approximately 90\% of MW-mass halos have had a central merger with a halo of
$\macc \ge 0.03\,\mviro$ since $z=2$.  If such mergers destroy galactic disks,
it is extremely difficult to understand the frequency of disk galaxies in
MW-mass halos ($\approx 70\%$; \citealt{weinmann2006, van-den-bosch2007,
  park2007}).  Around half of MW-mass halos have experienced a merger of $\macc
\ge 0.1 \,\mviro$ since $z=2$.  These events are likely to strongly affect
galaxy disks; however, the frequency with which they occur, coupled with the
possibility that a non-negligible fraction of such mergers may have sufficient
gas to reform a disk in the merger remnant, does not seem high enough to
present a major obstacle to forming galactic disks in a majority of MW-mass
halos.  It is therefore essential to simulate a large number of mergers in the
mass range of $0.03 \la \macc/\mviro \la 0.1$ with a distribution of realistic
orbits and galaxy models, including a variety of gas fractions.  This will
significantly advance our knowledge of disk formation and survivability in the
\lcdm\ cosmology.

\section{Discussion}
\label{sec:discussion}
\subsection{How (a)typical are the Aquarius halos?}
In general, the agreement between the properties of the Aquarius halos and
those of the full sample of MW-mass halos from the \millen\ is quite good: no
halo is an outlier from all of the relations studied here. Halos A and C form
somewhat earlier than average.  Halo C also has a larger concentration than is
typical.  On the other hand, neither A nor C seems to be an outlier in the
distribution of the spin parameter $\lambda^{\prime}$.  Halo F is somewhat
unusual in that it has a recent major merger; on the other hand, it is quite
typical in most of its other properties.  The Aquarius halos seem to reflect
the diverse properties of Milky Way-mass halos, both in
terms of assembly history and $z=0$ structure.

The Aquarius halos do not, however, seem to uniformly sample the spin parameter
$\lambda^{\prime}$: five of the six halos lie below the median of the
distribution, while the sixth is a merger remnant and may well relax to a lower
value as well.  We have explicitly checked that the isolation criterion in the
Aquarius halo selection does not bias the spin parameters, as the distribution
of $\lambda^{\prime}$ from the isolated and non-isolated samples are
statistically identical.  The somewhat low spin parameter distribution of the
Aquarius halos appears to be a statistical fluctuation.  This may be connected
to the relatively quiet recent merger history of the central subhalos for the
Aquarius halos.

\subsection{How (a)typical is the Galaxy's halo?}
While the Milky Way is frequently considered as the prototypical massive spiral
galaxy, this has its roots as much in convenience as in evidence.  While it is
difficult to use dissipationless $N$-body simulations to discern whether the MW
is indeed a typical galaxy for a halo of its mass, we can investigate whether
the Galaxy's halo itself is typical among those of similar mass.  Many possible
constraints are predicated on a more precise determination of the mass of the
MW's halo, however.  Current estimates vary by a factor of approximately 2 to 3
(with $1 \la M_{\rm vir, MW} \la 3\times 10^{12}\,\msun$).  This uncertainty
has a large effect on the probability of hosting massive satellite galaxies
such as the SMC and LMC: the chances of having a subhalo capable of hosting the
LMC or two subhalos capable of hosting the LMC and SMC are approximately
3-8\% for a MW mass of $10^{12}\,\msun$ but rise to 20-27\% if the
MW's halo has a mass of $2.5 \times 10^{12}\,\msun$.

On the other hand, a more massive Galactic halo implies that merger events of a
given $\macc/\mviro$ correspond to a larger fraction of the MW's disk mass.  If
the MW's thin disk must be explained via an unusually quiescent merger history
over the past 10 Gyr (e.g., \citealt{hammer2007}), we find that the MW's halo
must be even more unusual (in terms of merger history) if the halo is massive.
If we assume the MW cannot have merged with a $\macc > 3 M_{\rm disk, MW}$ halo
since $z=2$, this corresponds to $\mu=0.1$ for $M_{\rm MW}=10^{12}\,\msun$ and
$\mu=0.04$ if $M_{\rm MW}=2.5 \times 10^{12}\,\msun$.  From
Figure~\ref{fig:mergerHistory}, we find a 50\% chance for the former case but
only a 20\% chance for the latter.  If halos with $\macc=M_{\rm disk, MW}$
destroy thin disks, the probabilities drop to 12\% and 4\%, respectively.

A possible resolution of this tension -- that the existence of the LMC and SMC
argue for a more massive MW halo while the lack of recent mergers argues for a
less massive Galactic halo -- is to directly connect the two issues.  A recent
analysis of the LMC's orbital history \citep{besla2007} based on the updated
proper motions of the LMC \citep{kallivayalil2006} indicates that the LMC
likely fell into the Milky Way's halo within the last $\sim 3$ Gyr.  The LMC's
infall mass is likely to be in excess of $10^{11}\,\hmsun$ (and the SMC's
infall mass is likely to be only slightly less massive; \citealt{guo2010}),
meaning that the Milky Way has actually experienced a recent 1:10 merger at the
halo level.  \citet{stewart2008} show that Milky Way-mass halos typically
undergo one such merger at the halo level over the past 10 Gyr, perhaps
indicating that the Milky Way's accretion history is special only insofar as
the LMC fell in so recently.

\subsection{Cosmological Parameter Dependence}
\label{subsec:cosmo_params}
As noted in Sec.~\ref{subsec:msII}, the cosmological parameters used for the
\millen\ are slightly different from the most recent observational estimates:
the values of $\sigma_8=0.9$ and $n_s=1.0$ are approximately $3\,\sigma$ away
from the best-fitting values of 0.81 and 0.96, respectively.  Many of our
results are insensitive to such changes.  The abundance of MW-mass halos
differs by only $\sim 7\%$ in the WMAP5 cosmology compared to the
Millennium-II cosmology, for example, and the basic properties of halo assembly
histories scale weakly with $\sigma_8$ \citep{van-den-bosch2002}.  The subhalo
abundance of MW-mass halos could potentially be affected by $\sigma_8$ and
$n_s$, as decreasing $\sigma_8$ tends to reduce halo and subhalo concentrations
at a fixed mass (e.g., \citealt{maccio2008}).  This would tend to decrease the
amplitude of $\cmf$ in lower $\sigma_8$ cosmologies, as subhalos would be more
easily disrupted.  On the other hand, halos tend to form later in such
cosmologies, meaning that subhalos are subject to less dynamical evolution (see
the discussion at the start of Sec.~\ref{subsec:mass_func}); this would tend to
increase the amplitude of $\cmf$ for a lower $\sigma_8$.  These two effects
compensate to some extent, reducing the overall shift.

In order to properly ascertain the effects of cosmology on the structural
properties of MW-mass halos, it is necessary to run simulations that differ
only in their cosmological parameters and to search for differences in
statistical properties.  Nevertheless, there is reason to think that changing
from the cosmology of the Millennium, Millennium-II, and Aquarius simulations
to the WMAP 5 cosmology will not have a strong impact.  For example, the
difference in amplitude of $N(>\!\vmax/V_{\rm max, host})$ between
$\sigma_8=0.74$ and $1.0$ is only 14\% (\citealt{reed2005, diemand2008,
  diemand2009}), so reducing $\sigma_8$ from 0.9 to 0.81 will likely affect the
amplitude of the subhalo mass function by a few percent only.

\section{Conclusions}
\label{sec:conclusions}
We have presented a statistical study of Milky Way-mass halos from the
Millennium-II Simulation, which contains over 7000 halos in the mass range
$10^{11.5} \le \mvir \le 10^{12.5}\,\hmsun$.  Our principal results can be
summarized as follows:
\begin{itemize}
\item As several previous studies have shown, halo growth proceeds in an
  ``inside-out'' fashion: the central gravitational potential of a typical
  MW-mass halo reached half of its present-day value by $z=4$, on average,
  whereas the virial mass reached half of its present-day value at $z=1.2$.
\item The ratio of $z=0$ mass to the host halo mass for the most massive
  subhalo of a MW-mass halo has a broad distribution that peaks at 1\%.  The
  corresponding quantity computed using infall mass $\macc$ instead of $\msubo$
  peaks at 3.5\%
\item The differential and cumulative subhalo mass functions, computed in terms
  of $\mu \equiv \msub/\mvir$, are both
  well-fitted by a power law with an exponential cut-off at large $\mu$.  We
  find both the abundance per log decade in $\mu$ and the number
  of subhalos with mass greater than $\mu$ to be proportional to $\mu^{-0.935}$
  for small $\mu$.
\item The scatter in the cumulative mass function at fixed $\mu$ is
  only well-approximated by a Poisson distribution at large $\mu$
  (corresponding to mean occupation number $\langle N \rangle \la 4-5$).
  Intrinsic scatter of approximately 18\%, independent of the host halo mass,
  becomes increasingly important at lower $\mu$ (larger $\langle N \rangle$)
  and is dominant over Poisson scatter at $\mu \la 3\times 10^{-4}$ ($\langle N
  \rangle \ga 20$).
\item The Negative Binomial distribution with variance in $\cmf$ that is equal
  to the sum of a Poisson term and a fractional intrinsic scatter of 18\%
  matches the subhalo occupation data well at all $\mu$ for host halo masses
  between $10^{12}$ and $10^{15} \,\hmsun$.
\item The statistic $\alpha_2(\mu)$ [equation~(\ref{eq:alpha})], which is frequently
  used to characterize deviations of the HOD from Poisson, is not
  discriminating: distributions that differ strongly from Poisson can lead to
  few percent differences of $\alpha_2$ from unity.
\item Accretion redshifts of $z=0$ subhalos vary systematically with host halo
  mass and with the ratio of subhalo to host masses.  Massive subhalos at $z=0$ were
  typically accreted more recently.  This is a dynamical effect: massive
  subhalos accreted at early times either merge via dynamical friction or are heavily tidally
  truncated.  At fixed $\mu$, subhalos of more massive host halos were accreted
  more recently, reflecting the hierarchical build-up of dark matter halos.
  Subhalos with $\mu=0.001$ have a median accretion redshift of $\zacc=1.1$ for
  hosts with $\mvir \approx 10^{12}\,\hmsun$ and $\zacc=0.6$ for halos with
  $\mvir > 10^{13.5}\,\msun$.
\item The frequency of central mergers -- that is, the merger of a satellite
  with the central region of a MW-mass halo -- is a strong function of the
  satellite's $\macc$.  Since $z=2$ 90\% of MW-mass halos have experienced a
  central merger with a satellite having $\macc/\mviro > 0.03$ while 50\% have
  had such a merger with a $\macc/\mviro > 0.1$ satellite, and only 30\% with a
  $\macc/\mviro > 0.15$ satellite.  The compatibility of thin stellar disks
  with the frequent mergers expected in \lcdm\ thus depends strongly on exactly
  which mergers destroy disks.  If $\macc/\mviro > 0.03$ events are sufficient,
  the Milky Way must have an accretion history that lies among the quietest
  10\% for halos of similar mass, while if $\macc/\mviro > 0.1$ is required,
  then the Milky Way may have had a fairly typical merger history.
\end{itemize}

\section*{Acknowledgments}
We thank Andrey Kravtsov for stimulating discussions and the anonymous referee
for comments that improved the presentation of the paper.  We also thank Gerard
Lemson for his tireless efforts with the Millennium databases and for his
assistance in utilizing them efficiently.  The Millennium and Millennium-II
Simulation databases used in this paper and the web application providing
online access to them were constructed as part of the activities of the German
Astrophysical Virtual Observatory.  This work made extensive use of NASA's
Astrophysics Data System and of the astro-ph archive at arxiv.org.

\bibliography{draft}

\begin{thebibliography}{118}
\expandafter\ifx\csname natexlab\endcsname\relax\def\natexlab#1{#1}\fi

\bibitem[{{Angulo} {et~al.}(2009){Angulo}, {Lacey}, {Baugh}, \&
  {Frenk}}]{angulo2009}
{Angulo}, R.~E., {Lacey}, C.~G., {Baugh}, C.~M., \& {Frenk}, C.~S. 2009,
  \mnras, 399, 983

\bibitem[{{Bailin} \& {Steinmetz}(2005)}]{bailin2005}
{Bailin}, J., \& {Steinmetz}, M. 2005, \apj, 627, 647

\bibitem[{{Barnes} \& {Efstathiou}(1987)}]{barnes1987}
{Barnes}, J., \& {Efstathiou}, G. 1987, \apj, 319, 575

\bibitem[{{Battaglia} {et~al.}(2005){Battaglia}, {Helmi}, {Morrison},
  {Harding}, {Olszewski}, {Mateo}, {Freeman}, {Norris}, \&
  {Shectman}}]{battaglia2005}
{Battaglia}, G. {et~al.} 2005, \mnras, 364, 433

\bibitem[{{Belokurov} {et~al.}(2007){Belokurov}, {Zucker}, {Evans}, {Kleyna},
  {Koposov}, {Hodgkin}, {Irwin}, {Gilmore}, {Wilkinson}, {Fellhauer},
  {Bramich}, {Hewett}, {Vidrih}, {De Jong}, {Smith}, {Rix}, {Bell}, {Wyse},
  {Newberg}, {Mayeur}, {Yanny}, {Rockosi}, {Gnedin}, {Schneider}, {Beers},
  {Barentine}, {Brewington}, {Brinkmann}, {Harvanek}, {Kleinman}, {Krzesinski},
  {Long}, {Nitta}, \& {Snedden}}]{belokurov2007}
{Belokurov}, V. {et~al.} 2007, \apj, 654, 897

\bibitem[{{Benson} {et~al.}(2004){Benson}, {Lacey}, {Frenk}, {Baugh}, \&
  {Cole}}]{benson2004}
{Benson}, A.~J., {Lacey}, C.~G., {Frenk}, C.~S., {Baugh}, C.~M., \& {Cole}, S.
  2004, \mnras, 351, 1215

\bibitem[{{Berlind} \& {Weinberg}(2002)}]{berlind2002}
{Berlind}, A.~A., \& {Weinberg}, D.~H. 2002, \apj, 575, 587

\bibitem[{{Besla} {et~al.}(2007){Besla}, {Kallivayalil}, {Hernquist},
  {Robertson}, {Cox}, {van der Marel}, \& {Alcock}}]{besla2007}
{Besla}, G., {Kallivayalil}, N., {Hernquist}, L., {Robertson}, B., {Cox},
  T.~J., {van der Marel}, R.~P., \& {Alcock}, C. 2007, \apj, 668, 949

\bibitem[{{Bett} {et~al.}(2007){Bett}, {Eke}, {Frenk}, {Jenkins}, {Helly}, \&
  {Navarro}}]{bett2007}
{Bett}, P., {Eke}, V., {Frenk}, C.~S., {Jenkins}, A., {Helly}, J., \&
  {Navarro}, J. 2007, \mnras, 376, 215

\bibitem[{{Boylan-Kolchin} {et~al.}(2008){Boylan-Kolchin}, {Ma}, \&
  {Quataert}}]{boylan-kolchin2008}
{Boylan-Kolchin}, M., {Ma}, C.-P., \& {Quataert}, E. 2008, \mnras, 383, 93

\bibitem[{{Boylan-Kolchin} {et~al.}(2009){Boylan-Kolchin}, {Springel}, {White},
  {Jenkins}, \& {Lemson}}]{boylan-kolchin2009}
{Boylan-Kolchin}, M., {Springel}, V., {White}, S.~D.~M., {Jenkins}, A., \&
  {Lemson}, G. 2009, \mnras, 398, 1150

\bibitem[{{Brown} {et~al.}(2008){Brown}, {Zheng}, {White}, {Dey}, {Jannuzi},
  {Benson}, {Brand}, {Brodwin}, \& {Croton}}]{brown2008}
{Brown}, M.~J.~I. {et~al.} 2008, \apj, 682, 937

\bibitem[{{Bryan} \& {Norman}(1998)}]{bryan1998}
{Bryan}, G.~L., \& {Norman}, M.~L. 1998, \apj, 495, 80

\bibitem[{{Bullock} {et~al.}(2001{\natexlab{a}}){Bullock}, {Dekel}, {Kolatt},
  {Kravtsov}, {Klypin}, {Porciani}, \& {Primack}}]{bullock2001b}
{Bullock}, J.~S., {Dekel}, A., {Kolatt}, T.~S., {Kravtsov}, A.~V., {Klypin},
  A.~A., {Porciani}, C., \& {Primack}, J.~R. 2001{\natexlab{a}}, \apj, 555, 240

\bibitem[{{Bullock} {et~al.}(2001{\natexlab{b}}){Bullock}, {Kolatt}, {Sigad},
  {Somerville}, {Kravtsov}, {Klypin}, {Primack}, \& {Dekel}}]{bullock2001}
{Bullock}, J.~S., {Kolatt}, T.~S., {Sigad}, Y., {Somerville}, R.~S.,
  {Kravtsov}, A.~V., {Klypin}, A.~A., {Primack}, J.~R., \& {Dekel}, A.
  2001{\natexlab{b}}, \mnras, 321, 559

\bibitem[{{Cohn} \& {White}(2005)}]{cohn2005}
{Cohn}, J.~D., \& {White}, M. 2005, Astroparticle Physics, 24, 316

\bibitem[{{Cole} {et~al.}(2008){Cole}, {Helly}, {Frenk}, \&
  {Parkinson}}]{cole2008}
{Cole}, S., {Helly}, J., {Frenk}, C.~S., \& {Parkinson}, H. 2008, \mnras, 383,
  546

\bibitem[{{Cole} \& {Lacey}(1996)}]{cole1996}
{Cole}, S., \& {Lacey}, C. 1996, \mnras, 281, 716

\bibitem[{{Colless} {et~al.}(2001){Colless}, {Saglia}, {Burstein}, {Davies},
  {McMahan}, \& {Wegner}}]{colless2001}
{Colless}, M., {Saglia}, R.~P., {Burstein}, D., {Davies}, R.~L., {McMahan},
  R.~K., \& {Wegner}, G. 2001, \mnras, 321, 277

\bibitem[{{Conroy} \& {Wechsler}(2009)}]{conroy2009}
{Conroy}, C., \& {Wechsler}, R.~H. 2009, \apj, 696, 620

\bibitem[{{Conroy} {et~al.}(2006){Conroy}, {Wechsler}, \&
  {Kravtsov}}]{conroy2006}
{Conroy}, C., {Wechsler}, R.~H., \& {Kravtsov}, A.~V. 2006, \apj, 647, 201

\bibitem[{{Cuesta} {et~al.}(2008){Cuesta}, {Prada}, {Klypin}, \&
  {Moles}}]{cuesta2008}
{Cuesta}, A.~J., {Prada}, F., {Klypin}, A., \& {Moles}, M. 2008, \mnras, 389,
  385

\bibitem[{{De Lucia} {et~al.}(2004){De Lucia}, {Kauffmann}, {Springel},
  {White}, {Lanzoni}, {Stoehr}, {Tormen}, \& {Yoshida}}]{de-lucia2004}
{De Lucia}, G., {Kauffmann}, G., {Springel}, V., {White}, S.~D.~M., {Lanzoni},
  B., {Stoehr}, F., {Tormen}, G., \& {Yoshida}, N. 2004, \mnras, 348, 333

\bibitem[{{Dehnen} {et~al.}(2006){Dehnen}, {McLaughlin}, \&
  {Sachania}}]{dehnen2006}
{Dehnen}, W., {McLaughlin}, D.~E., \& {Sachania}, J. 2006, \mnras, 369, 1688

\bibitem[{{Diemand} {et~al.}(2007){Diemand}, {Kuhlen}, \&
  {Madau}}]{diemand2007a}
{Diemand}, J., {Kuhlen}, M., \& {Madau}, P. 2007, \apj, 667, 859

\bibitem[{{Diemand} {et~al.}(2008){Diemand}, {Kuhlen}, {Madau}, {Zemp},
  {Moore}, {Potter}, \& {Stadel}}]{diemand2008}
{Diemand}, J., {Kuhlen}, M., {Madau}, P., {Zemp}, M., {Moore}, B., {Potter},
  D., \& {Stadel}, J. 2008, \nat, 454, 735

\bibitem[{{Diemand} \& {Moore}(2009)}]{diemand2009}
{Diemand}, J., \& {Moore}, B. 2009, {arXiv:0906.4340 [astro-ph]}

\bibitem[{{D'Onghia} {et~al.}(2010){D'Onghia}, {Springel}, {Hernquist}, \&
  {Keres}}]{donghia2010}
{D'Onghia}, E., {Springel}, V., {Hernquist}, L., \& {Keres}, D. 2010, \apj,
  709, 1138

\bibitem[{{Dutton} {et~al.}(2007){Dutton}, {van den Bosch}, {Dekel}, \&
  {Courteau}}]{dutton2007}
{Dutton}, A.~A., {van den Bosch}, F.~C., {Dekel}, A., \& {Courteau}, S. 2007,
  \apj, 654, 27

\bibitem[{{Einasto}(1965)}]{einasto1965}
{Einasto}, J. 1965, Trudy Inst. Astrofiz. Alma-Ata, 51, 87

\bibitem[{{Fakhouri} \& {Ma}(2008)}]{fakhouri2008}
{Fakhouri}, O., \& {Ma}, C.-P. 2008, \mnras, 386, 577

\bibitem[{{Fukushige} \& {Makino}(2001)}]{fukushige2001}
{Fukushige}, T., \& {Makino}, J. 2001, \apj, 557, 533

\bibitem[{{Gao} {et~al.}(2008){Gao}, {Navarro}, {Cole}, {Frenk}, {White},
  {Springel}, {Jenkins}, \& {Neto}}]{gao2008}
{Gao}, L., {Navarro}, J.~F., {Cole}, S., {Frenk}, C.~S., {White}, S.~D.~M.,
  {Springel}, V., {Jenkins}, A., \& {Neto}, A.~F. 2008, \mnras, 387, 536

\bibitem[{{Gao} {et~al.}(2004){Gao}, {White}, {Jenkins}, {Stoehr}, \&
  {Springel}}]{gao2004b}
{Gao}, L., {White}, S.~D.~M., {Jenkins}, A., {Stoehr}, F., \& {Springel}, V.
  2004, \mnras, 355, 819

\bibitem[{{Genel} {et~al.}(2009){Genel}, {Genzel}, {Bouch{\'e}}, {Naab}, \&
  {Sternberg}}]{genel2009}
{Genel}, S., {Genzel}, R., {Bouch{\'e}}, N., {Naab}, T., \& {Sternberg}, A.
  2009, \apj, 701, 2002

\bibitem[{{Ghigna} {et~al.}(2000){Ghigna}, {Moore}, {Governato}, {Lake},
  {Quinn}, \& {Stadel}}]{ghigna2000}
{Ghigna}, S., {Moore}, B., {Governato}, F., {Lake}, G., {Quinn}, T., \&
  {Stadel}, J. 2000, \apj, 544, 616

\bibitem[{{Giocoli} {et~al.}(2008){Giocoli}, {Tormen}, \& {van den
  Bosch}}]{giocoli2008}
{Giocoli}, C., {Tormen}, G., \& {van den Bosch}, F.~C. 2008, \mnras, 386, 2135

\bibitem[{{Governato} {et~al.}(2009){Governato}, {Brook}, {Brooks}, {Mayer},
  {Willman}, {Jonsson}, {Stilp}, {Pope}, {Christensen}, {Wadsley}, \&
  {Quinn}}]{governato2009}
{Governato}, F. {et~al.} 2009, \mnras, 398, 312

\bibitem[{{Gunn} \& {Gott}(1972)}]{gunn1972}
{Gunn}, J.~E., \& {Gott}, J.~R.~I. 1972, \apj, 176, 1

\bibitem[{{Guo} {et~al.}(2010){Guo}, {White}, {Li}, \&
  {Boylan-Kolchin}}]{guo2010}
{Guo}, Q., {White}, S., {Li}, C., \& {Boylan-Kolchin}, M. 2010, \mnras, 404,
  1111

\bibitem[{{Guo} \& {White}(2008)}]{guo2008}
{Guo}, Q., \& {White}, S.~D.~M. 2008, \mnras, 384, 2

\bibitem[{{Hammer} {et~al.}(2007){Hammer}, {Puech}, {Chemin}, {Flores}, \&
  {Lehnert}}]{hammer2007}
{Hammer}, F., {Puech}, M., {Chemin}, L., {Flores}, H., \& {Lehnert}, M.~D.
  2007, \apj, 662, 322

\bibitem[{{Hayashi} {et~al.}(2003){Hayashi}, {Navarro}, {Taylor}, {Stadel}, \&
  {Quinn}}]{hayashi2003}
{Hayashi}, E., {Navarro}, J.~F., {Taylor}, J.~E., {Stadel}, J., \& {Quinn}, T.
  2003, \apj, 584, 541

\bibitem[{{Hopkins} {et~al.}(2009){Hopkins}, {Cox}, {Younger}, \&
  {Hernquist}}]{hopkins2009}
{Hopkins}, P.~F., {Cox}, T.~J., {Younger}, J.~D., \& {Hernquist}, L. 2009,
  \apj, 691, 1168

\bibitem[{{Ishiyama} {et~al.}(2009){Ishiyama}, {Fukushige}, \&
  {Makino}}]{ishiyama2009}
{Ishiyama}, T., {Fukushige}, T., \& {Makino}, J. 2009, \apj, 696, 2115

\bibitem[{{Jing}(2000)}]{jing2000}
{Jing}, Y.~P. 2000, \apj, 535, 30

\bibitem[{{Kallivayalil} {et~al.}(2006){Kallivayalil}, {van der Marel},
  {Alcock}, {Axelrod}, {Cook}, {Drake}, \& {Geha}}]{kallivayalil2006}
{Kallivayalil}, N., {van der Marel}, R.~P., {Alcock}, C., {Axelrod}, T.,
  {Cook}, K.~H., {Drake}, A.~J., \& {Geha}, M. 2006, \apj, 638, 772

\bibitem[{{Kazantzidis} {et~al.}(2008){Kazantzidis}, {Bullock}, {Zentner},
  {Kravtsov}, \& {Moustakas}}]{kazantzidis2008}
{Kazantzidis}, S., {Bullock}, J.~S., {Zentner}, A.~R., {Kravtsov}, A.~V., \&
  {Moustakas}, L.~A. 2008, \apj, 688, 254

\bibitem[{{Kazantzidis} {et~al.}(2009){Kazantzidis}, {Zentner}, {Kravtsov},
  {Bullock}, \& {Debattista}}]{kazantzidis2009}
{Kazantzidis}, S., {Zentner}, A.~R., {Kravtsov}, A.~V., {Bullock}, J.~S., \&
  {Debattista}, V.~P. 2009, \apj, 700, 1896

\bibitem[{{Kim} {et~al.}(1998){Kim}, {Staveley-Smith}, {Dopita}, {Freeman},
  {Sault}, {Kesteven}, \& {McConnell}}]{kim1998}
{Kim}, S., {Staveley-Smith}, L., {Dopita}, M.~A., {Freeman}, K.~C., {Sault},
  R.~J., {Kesteven}, M.~J., \& {McConnell}, D. 1998, \apj, 503, 674

\bibitem[{{Klypin} {et~al.}(1999){Klypin}, {Kravtsov}, {Valenzuela}, \&
  {Prada}}]{klypin1999}
{Klypin}, A., {Kravtsov}, A.~V., {Valenzuela}, O., \& {Prada}, F. 1999, \apj,
  522, 82

\bibitem[{{Klypin} {et~al.}(2002){Klypin}, {Zhao}, \&
  {Somerville}}]{klypin2002}
{Klypin}, A., {Zhao}, H., \& {Somerville}, R.~S. 2002, \apj, 573, 597

\bibitem[{{Komatsu} {et~al.}(2009){Komatsu}, {Dunkley}, {Nolta}, {Bennett},
  {Gold}, {Hinshaw}, {Jarosik}, {Larson}, {Limon}, {Page}, {Spergel},
  {Halpern}, {Hill}, {Kogut}, {Meyer}, {Tucker}, {Weiland}, {Wollack}, \&
  {Wright}}]{komatsu2009}
{Komatsu}, E. {et~al.} 2009, \apjs, 180, 330

\bibitem[{{Kravtsov}(2010)}]{kravtsov2010}
{Kravtsov}, A. 2010, Advances in Astronomy, 2010, 8

\bibitem[{{Kravtsov} {et~al.}(2004{\natexlab{a}}){Kravtsov}, {Berlind},
  {Wechsler}, {Klypin}, {Gottl{\"o}ber}, {Allgood}, \&
  {Primack}}]{kravtsov2004a}
{Kravtsov}, A.~V., {Berlind}, A.~A., {Wechsler}, R.~H., {Klypin}, A.~A.,
  {Gottl{\"o}ber}, S., {Allgood}, B., \& {Primack}, J.~R. 2004{\natexlab{a}},
  \apj, 609, 35

\bibitem[{{Kravtsov} {et~al.}(2004{\natexlab{b}}){Kravtsov}, {Gnedin}, \&
  {Klypin}}]{kravtsov2004}
{Kravtsov}, A.~V., {Gnedin}, O.~Y., \& {Klypin}, A.~A. 2004{\natexlab{b}},
  \apj, 609, 482

\bibitem[{{Lacey} \& {Cole}(1993)}]{lacey1993}
{Lacey}, C., \& {Cole}, S. 1993, \mnras, 262, 627

\bibitem[{{Lacey} \& {Cole}(1994)}]{lacey1994}
---. 1994, \mnras, 271, 676

\bibitem[{{Lemson} \& {Kauffmann}(1999)}]{lemson1999}
{Lemson}, G., \& {Kauffmann}, G. 1999, \mnras, 302, 111

\bibitem[{{Li} {et~al.}(2008){Li}, {Mo}, \& {Gao}}]{li2008}
{Li}, Y., {Mo}, H.~J., \& {Gao}, L. 2008, \mnras, 389, 1419

\bibitem[{{Li} \& {White}(2008)}]{li2008a}
{Li}, Y.-S., \& {White}, S.~D.~M. 2008, \mnras, 384, 1459

\bibitem[{{Loeb} \& {Peebles}(2003)}]{loeb2003}
{Loeb}, A., \& {Peebles}, P.~J.~E. 2003, \apj, 589, 29

\bibitem[{{Ma} \& {Fry}(2000)}]{ma2000}
{Ma}, C.-P., \& {Fry}, J.~N. 2000, \apj, 543, 503

\bibitem[{{Macci{\`o}} {et~al.}(2008){Macci{\`o}}, {Dutton}, \& {van den
  Bosch}}]{maccio2008}
{Macci{\`o}}, A.~V., {Dutton}, A.~A., \& {van den Bosch}, F.~C. 2008, \mnras,
  391, 1940

\bibitem[{{Macci{\`o}} {et~al.}(2007){Macci{\`o}}, {Dutton}, {van den Bosch},
  {Moore}, {Potter}, \& {Stadel}}]{maccio2007}
{Macci{\`o}}, A.~V., {Dutton}, A.~A., {van den Bosch}, F.~C., {Moore}, B.,
  {Potter}, D., \& {Stadel}, J. 2007, \mnras, 378, 55

\bibitem[{{Maller} {et~al.}(2002){Maller}, {Dekel}, \&
  {Somerville}}]{maller2002}
{Maller}, A.~H., {Dekel}, A., \& {Somerville}, R. 2002, \mnras, 329, 423

\bibitem[{{Mandelbaum} {et~al.}(2006){Mandelbaum}, {Seljak}, {Cool}, {Blanton},
  {Hirata}, \& {Brinkmann}}]{mandelbaum2006b}
{Mandelbaum}, R., {Seljak}, U., {Cool}, R.~J., {Blanton}, M., {Hirata}, C.~M.,
  \& {Brinkmann}, J. 2006, \mnras, 372, 758

\bibitem[{{McBride} {et~al.}(2009){McBride}, {Fakhouri}, \& {Ma}}]{mcbride2009}
{McBride}, J., {Fakhouri}, O., \& {Ma}, C.-P. 2009, \mnras, 398, 1858

\bibitem[{{Mo} {et~al.}(1998){Mo}, {Mao}, \& {White}}]{mo1998}
{Mo}, H.~J., {Mao}, S., \& {White}, S.~D.~M. 1998, \mnras, 295, 319

\bibitem[{{Moore} {et~al.}(1999){Moore}, {Ghigna}, {Governato}, {Lake},
  {Quinn}, {Stadel}, \& {Tozzi}}]{moore1999}
{Moore}, B., {Ghigna}, S., {Governato}, F., {Lake}, G., {Quinn}, T., {Stadel},
  J., \& {Tozzi}, P. 1999, \apjl, 524, L19

\bibitem[{{More} {et~al.}(2009){More}, {van den Bosch}, {Cacciato}, {Mo},
  {Yang}, \& {Li}}]{more2009}
{More}, S., {van den Bosch}, F.~C., {Cacciato}, M., {Mo}, H.~J., {Yang}, X., \&
  {Li}, R. 2009, \mnras, 392, 801

\bibitem[{{Moster} {et~al.}(2010){Moster}, {Macci{\`o}}, {Somerville},
  {Johansson}, \& {Naab}}]{moster2010a}
{Moster}, B.~P., {Macci{\`o}}, A.~V., {Somerville}, R.~S., {Johansson}, P.~H.,
  \& {Naab}, T. 2010, \mnras, 403, 1009

\bibitem[{{Navarro} {et~al.}(1996){Navarro}, {Frenk}, \& {White}}]{navarro1996}
{Navarro}, J.~F., {Frenk}, C.~S., \& {White}, S.~D.~M. 1996, \apj, 462, 563

\bibitem[{{Navarro} {et~al.}(1997){Navarro}, {Frenk}, \& {White}}]{navarro1997}
---. 1997, \apj, 490, 493

\bibitem[{{Navarro} {et~al.}(2010){Navarro}, {Ludlow}, {Springel}, {Wang},
  {Vogelsberger}, {White}, {Jenkins}, {Frenk}, \& {Helmi}}]{navarro2010}
{Navarro}, J.~F. {et~al.} 2010, \mnras, 402, 21

\bibitem[{{Neistein} \& {Dekel}(2008)}]{neistein2008}
{Neistein}, E., \& {Dekel}, A. 2008, \mnras, 388, 1792

\bibitem[{{Neistein} {et~al.}(2006){Neistein}, {van den Bosch}, \&
  {Dekel}}]{neistein2006}
{Neistein}, E., {van den Bosch}, F.~C., \& {Dekel}, A. 2006, \mnras, 372, 933

\bibitem[{{Neto} {et~al.}(2007){Neto}, {Gao}, {Bett}, {Cole}, {Navarro},
  {Frenk}, {White}, {Springel}, \& {Jenkins}}]{neto2007}
{Neto}, A.~F. {et~al.} 2007, \mnras, 381, 1450

\bibitem[{{Park} {et~al.}(2007){Park}, {Choi}, {Vogeley}, {Gott}, \&
  {Blanton}}]{park2007}
{Park}, C., {Choi}, Y., {Vogeley}, M.~S., {Gott}, J.~R.~I., \& {Blanton}, M.~R.
  2007, \apj, 658, 898

\bibitem[{{Peacock} \& {Smith}(2000)}]{peacock2000}
{Peacock}, J.~A., \& {Smith}, R.~E. 2000, \mnras, 318, 1144

\bibitem[{{Press} \& {Schechter}(1974)}]{press1974}
{Press}, W.~H., \& {Schechter}, P. 1974, \apj, 187, 425

\bibitem[{{Purcell} {et~al.}(2009){Purcell}, {Kazantzidis}, \&
  {Bullock}}]{purcell2009}
{Purcell}, C.~W., {Kazantzidis}, S., \& {Bullock}, J.~S. 2009, \apjl, 694, L98

\bibitem[{{Reed} {et~al.}(2005){Reed}, {Governato}, {Quinn}, {Gardner},
  {Stadel}, \& {Lake}}]{reed2005}
{Reed}, D., {Governato}, F., {Quinn}, T., {Gardner}, J., {Stadel}, J., \&
  {Lake}, G. 2005, \mnras, 359, 1537

\bibitem[{{Sakamoto} {et~al.}(2003){Sakamoto}, {Chiba}, \&
  {Beers}}]{sakamoto2003}
{Sakamoto}, T., {Chiba}, M., \& {Beers}, T.~C. 2003, \aap, 397, 899

\bibitem[{{Scannapieco} {et~al.}(2009){Scannapieco}, {White}, {Springel}, \&
  {Tissera}}]{scannapieco2009}
{Scannapieco}, C., {White}, S.~D.~M., {Springel}, V., \& {Tissera}, P.~B. 2009,
  \mnras, 396, 696

\bibitem[{{Scoccimarro} {et~al.}(2001){Scoccimarro}, {Sheth}, {Hui}, \&
  {Jain}}]{scoccimarro2001}
{Scoccimarro}, R., {Sheth}, R.~K., {Hui}, L., \& {Jain}, B. 2001, \apj, 546, 20

\bibitem[{{Seljak}(2000)}]{seljak2000}
{Seljak}, U. 2000, \mnras, 318, 203

\bibitem[{{Spergel} {et~al.}(2003){Spergel}, {Verde}, {Peiris}, {Komatsu},
  {Nolta}, {Bennett}, {Halpern}, {Hinshaw}, {Jarosik}, {Kogut}, {Limon},
  {Meyer}, {Page}, {Tucker}, {Weiland}, {Wollack}, \& {Wright}}]{spergel2003}
{Spergel}, D.~N. {et~al.} 2003, \apjs, 148, 175

\bibitem[{{Springel} \& {Hernquist}(2005)}]{springel2005c}
{Springel}, V., \& {Hernquist}, L. 2005, \apjl, 622, L9

\bibitem[{{Springel} {et~al.}(2008){Springel}, {Wang}, {Vogelsberger},
  {Ludlow}, {Jenkins}, {Helmi}, {Navarro}, {Frenk}, \& {White}}]{springel2008}
{Springel}, V. {et~al.} 2008, \mnras, 391, 1685

\bibitem[{{Springel} {et~al.}(2005){Springel}, {White}, {Jenkins}, {Frenk},
  {Yoshida}, {Gao}, {Navarro}, {Thacker}, {Croton}, {Helly}, {Peacock}, {Cole},
  {Thomas}, {Couchman}, {Evrard}, {Colberg}, \& {Pearce}}]{springel2005b}
---. 2005, \nat, 435, 629

\bibitem[{{Springel} {et~al.}(2001){Springel}, {White}, {Tormen}, \&
  {Kauffmann}}]{springel2001a}
{Springel}, V., {White}, S.~D.~M., {Tormen}, G., \& {Kauffmann}, G. 2001,
  \mnras, 328, 726

\bibitem[{{Stadel} {et~al.}(2009){Stadel}, {Potter}, {Moore}, {Diemand},
  {Madau}, {Zemp}, {Kuhlen}, \& {Quilis}}]{stadel2009}
{Stadel}, J., {Potter}, D., {Moore}, B., {Diemand}, J., {Madau}, P., {Zemp},
  M., {Kuhlen}, M., \& {Quilis}, V. 2009, \mnras, 398, L21

\bibitem[{{Stanimirovi{\'c}} {et~al.}(2004){Stanimirovi{\'c}},
  {Staveley-Smith}, \& {Jones}}]{stanimirovic2004}
{Stanimirovi{\'c}}, S., {Staveley-Smith}, L., \& {Jones}, P.~A. 2004, \apj,
  604, 176

\bibitem[{{Stewart} {et~al.}(2009){Stewart}, {Bullock}, {Wechsler}, \&
  {Maller}}]{stewart2009a}
{Stewart}, K.~R., {Bullock}, J.~S., {Wechsler}, R.~H., \& {Maller}, A.~H. 2009,
  \apj, 702, 307

\bibitem[{{Stewart} {et~al.}(2008){Stewart}, {Bullock}, {Wechsler}, {Maller},
  \& {Zentner}}]{stewart2008}
{Stewart}, K.~R., {Bullock}, J.~S., {Wechsler}, R.~H., {Maller}, A.~H., \&
  {Zentner}, A.~R. 2008, \apj, 683, 597

\bibitem[{{Tasitsiomi} {et~al.}(2004){Tasitsiomi}, {Kravtsov}, {Gottl{\"o}ber},
  \& {Klypin}}]{tasitsiomi2004}
{Tasitsiomi}, A., {Kravtsov}, A.~V., {Gottl{\"o}ber}, S., \& {Klypin}, A.~A.
  2004, \apj, 607, 125

\bibitem[{{Tollerud} {et~al.}(2008){Tollerud}, {Bullock}, {Strigari}, \&
  {Willman}}]{tollerud2008}
{Tollerud}, E.~J., {Bullock}, J.~S., {Strigari}, L.~E., \& {Willman}, B. 2008,
  \apj, 688, 277

\bibitem[{{Tormen} {et~al.}(1998){Tormen}, {Diaferio}, \& {Syer}}]{tormen1998}
{Tormen}, G., {Diaferio}, A., \& {Syer}, D. 1998, \mnras, 299, 728

\bibitem[{{Toth} \& {Ostriker}(1992)}]{toth1992}
{Toth}, G., \& {Ostriker}, J.~P. 1992, \apj, 389, 5

\bibitem[{{Vale} \& {Ostriker}(2004)}]{vale2004}
{Vale}, A., \& {Ostriker}, J.~P. 2004, \mnras, 353, 189

\bibitem[{{van den Bosch}(2002)}]{van-den-bosch2002}
{van den Bosch}, F.~C. 2002, \mnras, 331, 98

\bibitem[{{van den Bosch} {et~al.}(2007){van den Bosch}, {Yang}, {Mo},
  {Weinmann}, {Macci{\`o}}, {More}, {Cacciato}, {Skibba}, \&
  {Kang}}]{van-den-bosch2007}
{van den Bosch}, F.~C. {et~al.} 2007, \mnras, 376, 841

\bibitem[{{Velazquez} \& {White}(1999)}]{velazquez1999}
{Velazquez}, H., \& {White}, S.~D.~M. 1999, \mnras, 304, 254

\bibitem[{{Vitvitska} {et~al.}(2002){Vitvitska}, {Klypin}, {Kravtsov},
  {Wechsler}, {Primack}, \& {Bullock}}]{vitvitska2002}
{Vitvitska}, M., {Klypin}, A.~A., {Kravtsov}, A.~V., {Wechsler}, R.~H.,
  {Primack}, J.~R., \& {Bullock}, J.~S. 2002, \apj, 581, 799

\bibitem[{{Walsh} {et~al.}(2009){Walsh}, {Willman}, \& {Jerjen}}]{walsh2009}
{Walsh}, S.~M., {Willman}, B., \& {Jerjen}, H. 2009, \aj, 137, 450

\bibitem[{{Warren} {et~al.}(1992){Warren}, {Quinn}, {Salmon}, \&
  {Zurek}}]{warren1992}
{Warren}, M.~S., {Quinn}, P.~J., {Salmon}, J.~K., \& {Zurek}, W.~H. 1992, \apj,
  399, 405

\bibitem[{{Wechsler} {et~al.}(2002){Wechsler}, {Bullock}, {Primack},
  {Kravtsov}, \& {Dekel}}]{wechsler2002}
{Wechsler}, R.~H., {Bullock}, J.~S., {Primack}, J.~R., {Kravtsov}, A.~V., \&
  {Dekel}, A. 2002, \apj, 568, 52

\bibitem[{{Weinmann} {et~al.}(2006){Weinmann}, {van den Bosch}, {Yang}, \&
  {Mo}}]{weinmann2006}
{Weinmann}, S.~M., {van den Bosch}, F.~C., {Yang}, X., \& {Mo}, H.~J. 2006,
  \mnras, 366, 2

\bibitem[{{Wetzel} \& {White}(2010)}]{wetzel2010}
{Wetzel}, A.~R., \& {White}, M. 2010, \mnras, 403, 1072

\bibitem[{{White} \& {Rees}(1978)}]{white1978a}
{White}, S.~D.~M., \& {Rees}, M.~J. 1978, \mnras, 183, 341

\bibitem[{{Wilkinson} \& {Evans}(1999)}]{wilkinson1999}
{Wilkinson}, M.~I., \& {Evans}, N.~W. 1999, \mnras, 310, 645

\bibitem[{{Xue} {et~al.}(2008){Xue}, {Rix}, {Zhao}, {Re Fiorentin}, {Naab},
  {Steinmetz}, {van den Bosch}, {Beers}, {Lee}, {Bell}, {Rockosi}, {Yanny},
  {Newberg}, {Wilhelm}, {Kang}, {Smith}, \& {Schneider}}]{xue2008}
{Xue}, X.~X. {et~al.} 2008, \apj, 684, 1143

\bibitem[{{York} {et~al.}(2000){York}, {Adelman}, {Anderson}, {Anderson},
  {Annis}, {Bahcall}, {Bakken}, {Barkhouser}, {Bastian}, {Berman}, {Boroski},
  {Bracker}, {Briegel}, {Briggs}, {Brinkmann}, {Brunner}, {Burles}, {Carey},
  {Carr}, {Castander}, {Chen}, {Colestock}, {Connolly}, {Crocker}, {Csabai},
  {Czarapata}, {Davis}, {Doi}, {Dombeck}, {Eisenstein}, {Ellman}, {Elms},
  {Evans}, {Fan}, {Federwitz}, {Fiscelli}, {Friedman}, {Frieman}, {Fukugita},
  {Gillespie}, {Gunn}, {Gurbani}, {de Haas}, {Haldeman}, {Harris}, {Hayes},
  {Heckman}, {Hennessy}, {Hindsley}, {Holm}, {Holmgren}, {Huang}, {Hull},
  {Husby}, {Ichikawa}, {Ichikawa}, {Ivezi{\'c}}, {Kent}, {Kim}, {Kinney},
  {Klaene}, {Kleinman}, {Kleinman}, {Knapp}, {Korienek}, {Kron}, {Kunszt},
  {Lamb}, {Lee}, {Leger}, {Limmongkol}, {Lindenmeyer}, {Long}, {Loomis},
  {Loveday}, {Lucinio}, {Lupton}, {MacKinnon}, {Mannery}, {Mantsch}, {Margon},
  {McGehee}, {McKay}, {Meiksin}, {Merelli}, {Monet}, {Munn}, {Narayanan},
  {Nash}, {Neilsen}, {Neswold}, {Newberg}, {Nichol}, {Nicinski}, {Nonino},
  {Okada}, {Okamura}, {Ostriker}, {Owen}, {Pauls}, {Peoples}, {Peterson},
  {Petravick}, {Pier}, {Pope}, {Pordes}, {Prosapio}, {Rechenmacher}, {Quinn},
  {Richards}, {Richmond}, {Rivetta}, {Rockosi}, {Ruthmansdorfer}, {Sandford},
  {Schlegel}, {Schneider}, {Sekiguchi}, {Sergey}, {Shimasaku}, {Siegmund},
  {Smee}, {Smith}, {Snedden}, {Stone}, {Stoughton}, {Strauss}, {Stubbs},
  {SubbaRao}, {Szalay}, {Szapudi}, {Szokoly}, {Thakar}, {Tremonti}, {Tucker},
  {Uomoto}, {Vanden Berk}, {Vogeley}, {Waddell}, {Wang}, {Watanabe},
  {Weinberg}, {Yanny}, \& {Yasuda}}]{york2000}
{York}, D.~G. {et~al.} 2000, \aj, 120, 1579

\bibitem[{{Zhao} {et~al.}(2003{\natexlab{a}}){Zhao}, {Jing}, {Mo}, \&
  {B{\"o}rner}}]{zhao2003a}
{Zhao}, D.~H., {Jing}, Y.~P., {Mo}, H.~J., \& {B{\"o}rner}, G.
  2003{\natexlab{a}}, \apjl, 597, L9

\bibitem[{{Zhao} {et~al.}(2009){Zhao}, {Jing}, {Mo}, \&
  {B{\"o}rner}}]{zhao2009}
---. 2009, \apj, 707, 354

\bibitem[{{Zhao} {et~al.}(2003{\natexlab{b}}){Zhao}, {Mo}, {Jing}, \&
  {B{\"o}rner}}]{zhao2003}
{Zhao}, D.~H., {Mo}, H.~J., {Jing}, Y.~P., \& {B{\"o}rner}, G.
  2003{\natexlab{b}}, \mnras, 339, 12

\bibitem[{{Zheng} {et~al.}(2005){Zheng}, {Berlind}, {Weinberg}, {Benson},
  {Baugh}, {Cole}, {Dav{\'e}}, {Frenk}, {Katz}, \& {Lacey}}]{zheng2005}
{Zheng}, Z. {et~al.} 2005, \apj, 633, 791

\end{thebibliography}
\appendix
\section{Subhalo masses and circular velocities}

\begin{figure}
 \centering
 \includegraphics[scale=0.56, viewport=0 0 421 415, clip]{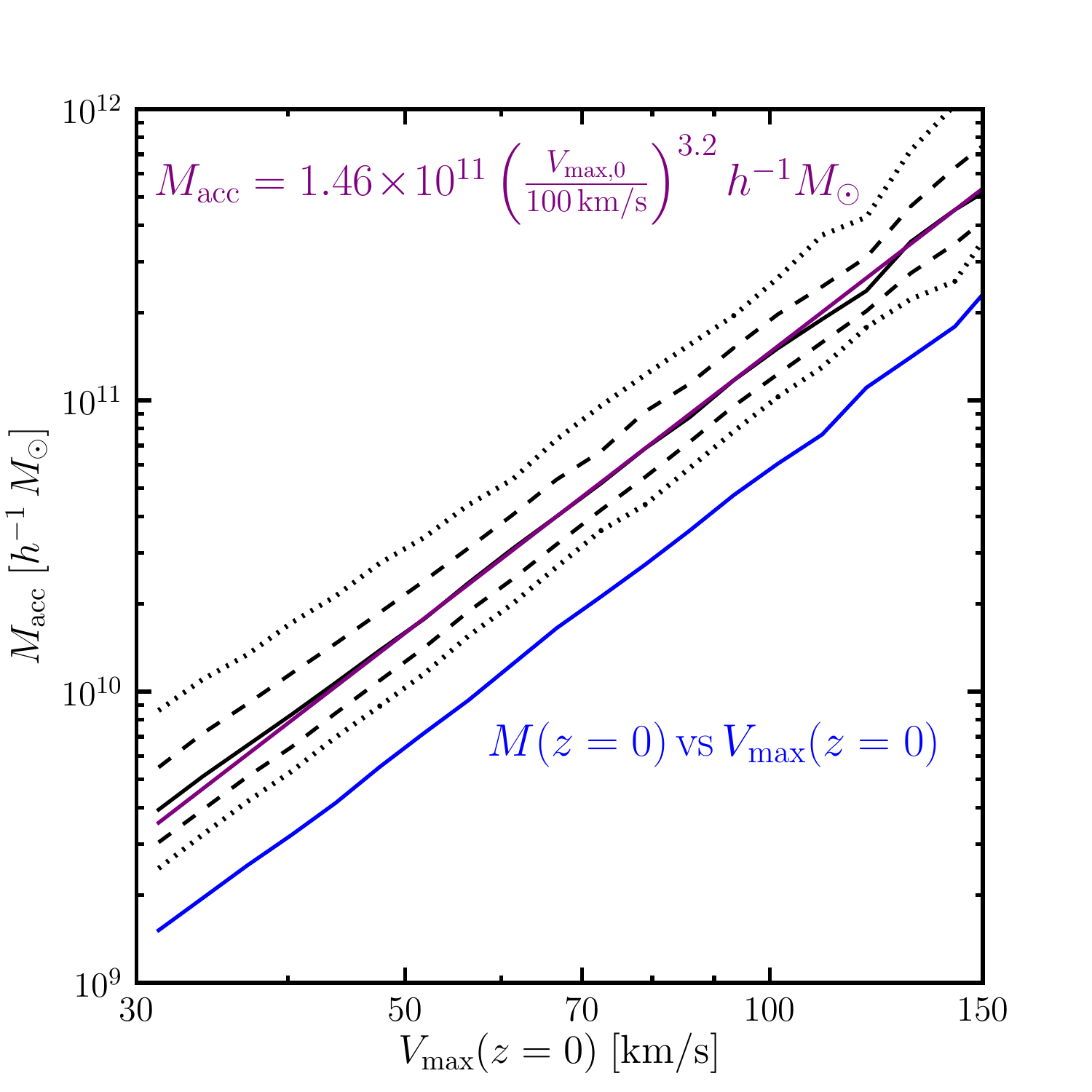}
 \caption{
   $\macc$ as a function of $\vmax$ for subhalos surviving to $z=0$.
   The solid curve shows the median $\macc$, while the dashed (dotted) curves
   show the 25 and 75 (10 and 90) percentiles.  Also plotted, in magenta, is
   the best-fitting power law relation between $\macc$ and $\vmaxo$, as
   well as the median relation between $M(z=0)$ and $\vmaxo$ for subhalos
   (blue curve).  The virtually constant offset between the blue and solid black
   curves shows that the typical surviving subhalo loses a factor of $\sim 2.5$ in mass
   between infall and the present day, irrespective of $\vmaxo$.  
  \label{fig:z0vmax_vs_mmax}
}
\end{figure}

Figure~\ref{fig:z0vmax_vs_mmax} shows the median relations between $\macc$ and
$\vmaxo$ (solid black line) and between $\msubo$ and $\vmaxo$ (solid blue
line); the latter is identical to the one plotted in Fig.~\ref{fig:massVsVmax}.
Also plotted are the regions centered on the median $\macc-\vmaxo$ relation
containing 50 and 80\% of the distribution (dashed and dotted curves,
respectively).  The $\macc-\vmaxo$ relation follows the same slope as the
$\msubo-\vmaxo$ relation but has an amplitude that is approximately 2.4 times
larger: the solid magenta line shows the result of multiplying the fit to
$\msubo-\vmaxo$ at $z=0$ by a factor of 2.44.  The average subhalo at $z=0$ was
therefore approximately 2.4 times more massive at accretion, irrespective of
its present $\vmax$.  The spread in the $\macc-\vmaxo$ relation is
non-negligible: the 90\% value of $\macc$ is typically a factor of 2.5 larger
than the 10\% value at fixed $\vmax$.  10\% of halos have lost over 80\% of
their mass while another 10\% have lost less than 30\%.

\begin{figure}
 \centering
 \includegraphics[scale=0.56, viewport=0 0 421 415, clip]{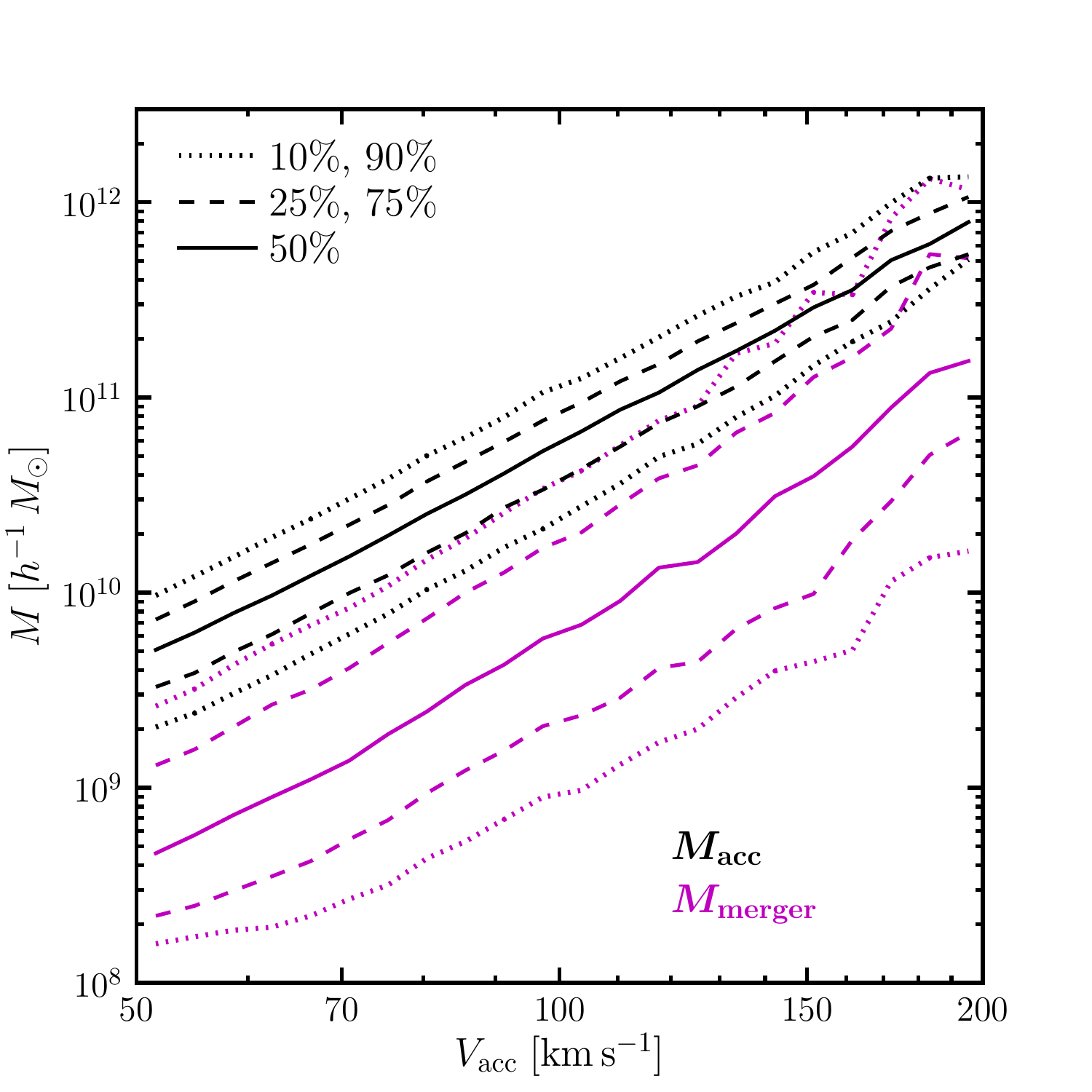}
 \caption{
   $\macc$ as a function of $\vmax$ for subhalos with $\vacc \ge 50 \,\kms$
   that merge with their host.
   The solid curve shows the median $\macc$, while the dashed (dotted) curves
   show the 25 and 75 (10 and 90) percentiles.  
   Also plotted, in magenta, is
   the relation between $\vmax$ and $\mmerge$, the mass of the subhalo
   immediately prior to merging.  The median $\vmax-\mmerge$ relation is
   nearly parallel to the median $\vmax-\macc$ relation but is lower in
   amplitude by approximately 1 dex, indicating that subhalos lose
   approximately 90\% of their mass before merging for all $\vacc$.
 \label{fig:mergerMass}
}
\end{figure}
Figure~\ref{fig:mergerMass} shows the median relation between $\macc$ and
$\vacc$ (solid black curve), along with the ranges containing 50 and 80\% of
the distribution (dashed and dotted black curves).  Our lower limit of $\vacc
=50\, \kms$ corresponds to a median $\macc$ of $5 \times 10^{9} \,\hmsun$.
Figure~\ref{fig:mergerMass} also shows in magenta the relation between $\vacc$
and $\mmerge$, defined to be the mass of the subhalo at the snapshot
immediately prior to merging.  The relation is nearly parallel to
$\macc(\vacc)$ (the solid black line) but is lower in amplitude by
approximately a factor of 10 at all $\vacc$.  This means that these subhalos
lose approximately 90\% of their mass prior to merging, irrespective of their
mass at infall.

\label{lastpage}
\end{document}